\documentclass[aps,prb,twocolumn,superscriptaddress,longbibliography]{revtex4-2}

\usepackage{graphicx} 
\usepackage{physics}  
\usepackage{amsmath}  
\usepackage{array}    
\usepackage{subfigure} 
\usepackage[utf8]{inputenc} 
\usepackage{braket}
\usepackage{comment}
\usepackage{hyperref}
\usepackage{booktabs}
\hypersetup{
    colorlinks=true,        
    linkcolor=blue,         
    citecolor=blue,         
    urlcolor=blue           
}

\begin{document}

\title{Confinement, deconfinement, and bound states in the spin-$1$ and spin-$3/2$ generalizations of the Majumdar--Ghosh chain}

\author{Aman Sharma}
\email{a.sharma@epfl.ch}
\affiliation{Institute of Physics, \'Ecole Polytechnique F\'ed\'erale de Lausanne (EPFL), CH-1015 Lausanne, Switzerland\looseness=-1}

\author{Mithilesh Nayak}
\affiliation{Department of Physics, University of Fribourg, 1700 Fribourg, Switzerland}
\affiliation{Department of Physics and Astronomy, The University of Tennessee, Knoxville, Tennessee 37996, USA}

\author{Natalia Chepiga}

\affiliation{Kavli Institute of Nanoscience, Delft University of Technology, Lorentzweg 1, 2628 CJ Delft, The Netherlands}

\author{Henrik M. R{\o}nnow}
	
\affiliation{Institute of Physics, \'Ecole Polytechnique F\'ed\'erale de Lausanne (EPFL), CH-1015 Lausanne, Switzerland\looseness=-1}

\author{Fr\'ed\'eric Mila}
\affiliation{Institute of Physics, \'Ecole Polytechnique F\'ed\'erale de Lausanne (EPFL), CH-1015 Lausanne, Switzerland\looseness=-1}

\date{\today}

\begin{abstract}
We investigate the nature of low-energy excitations in a spin chain with antiferrmomagnetic nearest-neighbor $J_1$, next-nearest-neighbor $J_2$, and three-site $J_3$ interactions using the time-dependent density matrix renormalization group and the single mode approximation techniques. In the absence of the $J_2$ interaction, we identify clear distinctions in the spectral functions in the fully dimerized phase across the exactly dimerized line for different magnitudes of the spins. In contrast to the spin-$1/2$ chain, where the spinon continuum dominates the spectral functions, the magnon modes are prominent in the spectral functions of the spin-$1$ and spin-$3/2$ chains. Through single mode approximation and valence bond solid approaches, we disentangle magnon and spinon contributions to the spectral functions. After including the $J_2$ interactions, for the spin-$1$ chain we trace the evolution of the dynamical structure factor along the phase transition line between the Haldane phase and the fully dimerized phase. We find that the excitation spectrum is a continuum along this line and the spectral gap closes as the order of the transition changes from first order to second order. Along the line of first-order transitions, the spinon-like domain walls are deconfined, and the model exhibits their confinement into discrete bound states away from the transition line. A similar phenomenon occurs in the spin-$3/2$ chain across the phase transition between partially dimerized to fully dimerized phases, revealing a universal spinon confinement phenomenon across first-order phase transitions. This study presents the dynamical structure factor corresponding to the ground state phase diagram and establishes a unified quasiparticle framework for understanding the fundamental nature of excitations across distinct quantum phases in frustrated $J_1$-$J_2$-$J_3$ Heisenberg spin chains.

\end{abstract}

\maketitle

\section{Introduction}

Frustrated spin chains with competing interactions, such as next-nearest-neighbor couplings or multisite exchange terms, provide a rich setting for exploring subtle quantum phenomena, including spontaneous dimerization, gap formation, and quantum criticality~\cite{tonegawa1987ground, okamoto1992fluid, eggert1996numerical, giamarchi2003quantum, affleck1989quantum, lacroix2011introduction}. Theoretical studies of these systems offer valuable insights into the confinement and deconfinement of fractional excitations~\cite{sharma2025bound, vanderstraeten2020spinon,sorensen1998soliton}.

For spin-\(1/2\), the paradigmatic model is the $J_1$--$J_2$ chain, famous for its exactly dimerized ground state at the Majumdar--Ghosh point $J_2 = J_1/2$. There are two natural generalizations of the spin-\(1/2\) \(J_1\)-\(J_2\) chain for larger spins. One is the \( J_1 \)-\( J_2 \) model for arbitrary spin \( S \), which was explored in our earlier work~\cite{sharma3by2}. The other is the \( J_1 \)-\( J_2 \)-\( J_3 \) model for higher spins, which has an exact dimerized ground state along the generalized Majumdar-Ghosh (MG) line \cite{michaud2012antiferromagnetic,wang2013dimerizations}. In the present article, we focus on this latter generalization and investigate the physics around this MG line. We consider the frustrated Heisenberg spin chain with nearest-neighbor, next-nearest-neighbor, and three-site interactions, commonly referred to as the \( J_1 \)-\( J_2 \)-\( J_3 \) model:

\begin{eqnarray}
H &=& J_1 \sum_i \mathbf{S}_i \cdot \mathbf{S}_{i+1} + J_2 \sum_i \mathbf{S}_i \cdot \mathbf{S}_{i+2} \nonumber \\
  &&+ J_3 \sum_i [(\mathbf{S}_{i-1} \cdot \mathbf{S}_i)(\mathbf{S}_i \cdot \mathbf{S}_{i+1}) +  \text{h.c.}] 
\end{eqnarray}

We focus on the cases where the spin magnitude is \( S=1 \) and \( S=3/2 \), and study the model across various regimes of frustration. Notably, for \( S=1/2 \), the three-site term reduces to a next-nearest-neighbor interaction via the identity \( J_2 = J_3/2 \), rendering the \( J_1 \)-\( J_2 \)-\( J_3 \) model equivalent to the well-known \( J_1 \)-\( J_2 \) chain~\cite{michaud2012antiferromagnetic,michaud2013realization}.

The $J_1$-$J_2$ model for spin-1/2 undergoes a Kosterlitz-Thouless (KT) transition from a gapless critical phase to a spontaneously dimerized phase at \( J_2/J_1 \approx 0.2411 \)~\cite{okamoto1992fluid}, with an exact result at the MG point \( J_2/J_1 = 1/2 \) where the ground state is exactly dimerized as a product of singlet pairs on alternating bonds~\cite{bursill1995numerical,nomura2003onset,majumdar1969next}. For arbitrary spin $S$, Michaud et al.~\cite{michaud2012antiferromagnetic} showed that an exact dimerized ground state exists in the $J_1$--$J_3$ model when
\begin{equation}
\frac{J_3}{J_1} = \frac{1}{4S(S+1) - 2},
\label{eq:MGpointJ2zero}
\end{equation}
which generalizes the MG point ($J_2=0$) to higher spins. More generally, along the exactly-dimerized line including finite $J_2$ one has~\cite{wang2013dimerizations}
\begin{equation}
\label{eq:MGline}
\frac{J_3}{J_1-2J_2}=\frac{1}{4S(S+1)-2},
\end{equation}
which reduces to Eq.~\eqref{eq:MGpointJ2zero} when $J_2=0$.  In the case of the spin-1 \( J_1 \)-\( J_3 \) model (with \( J_2=0 \)), the system is in the Haldane phase for small \( J_3 \) \cite{haldane1983continuum,aklt1987rigorous,auerbach1998interacting}, and undergoes a continuous phase transition into a dimerized state at \( J_3/J_1 \approx 0.111 \)~\cite{michaud2012antiferromagnetic}, before the generalized MG point (\( J_3/J_1 = 1/6 \)). At this critical point, the low-energy physics is described by a Wess-Zumino-Witten (WZW) conformal field theory $SU(2)_2$ with central charge \( c=3/2 \) \cite{affleck1987critical}, the same critical theory as that realized in the integrable spin-1 chain with negative biquadratic interaction\cite{babujian1982exact,takhtajan1982picture}. In the \( J_1 \)-\( J_3 \) spin-3/2 model, for small \( J_3 \), the system resides in a critical phase~\cite{affleck1988valence, niggemann1997quantum, tu2008valence}, and increasing \( J_3 \) leads to a continuous transition into the fully dimerized phase \cite{chepiga2020floating}. At the MG point of the spin-$3/2$ chain ( $J_3/J_1 = 1/13$ ), the two ground states are exactly describable as products of singlets\cite{michaud2013realization,chepiga2020floating}.

\begin{figure*}[t]
    \centering

    \includegraphics[width=1\textwidth]{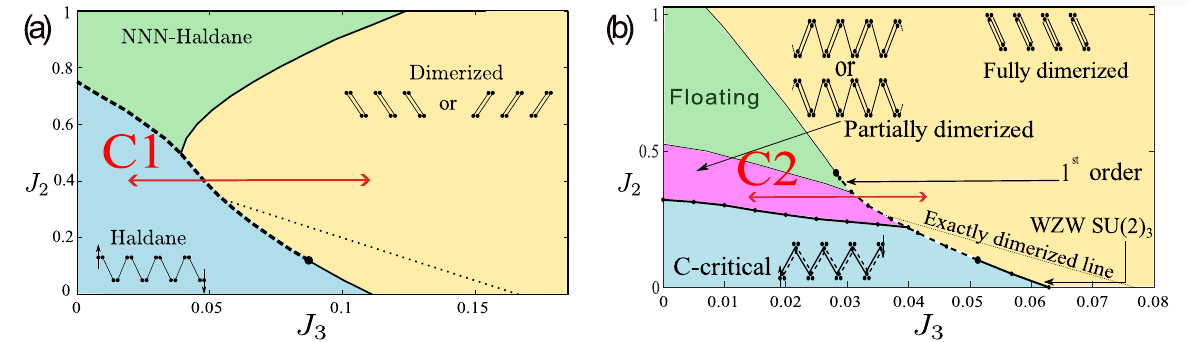}
            \label{}

    \caption{(a) Phase diagram of the spin-1 $J_1$–$J_2$–$J_3$ chain adapted from ref.~\cite{chepiga2016spontaneous}, showing transitions between Haldane, dimerized, and NNN-Haldane phases. (b) Phase diagram of the spin-$3/2$ $J_1$–$J_2$–$J_3$ chain adapted from ref.~\cite{chepiga2020floating}, featuring critical, partially dimerized, floating, and fully dimerized phases. In both panels, solid and dashed lines indicate continuous and first-order transitions, respectively. The fully dimerized state is exact along the dotted lines. Double-headed red arrows indicate the cuts (C1 and C2) along which dynamical calculations were performed to investigate the nature of excitations across the first-order transitions in the spin-1 and spin-\(3/2\) models, at \( J_2 = 0.4 J_1 \) and \( J_2 = 0.328 J_1 \), respectively.}
    \label{phasediags}
\end{figure*}

The  $J_1$-$J_2$-$J_3$ spin-$1$ chain exhibits a rich phase diagram comprising a Haldane phase, NNN Haldane and dimerized phase (Fig. \ref{phasediags}a) \cite{chepiga2016dimerization}. It has been established that the transition from the Haldane to the dimerized phase is second order for small $J_2$ and becomes first order as $J_2$ increases \cite{chepiga2016dimerization}. For spin-$3/2$, the model reveals even richer behavior (Fig. \ref{phasediags}b). The phase diagram includes a critical phase, characterized by gapless excitations and commensurate spin-spin correlations, a partially dimerized phase, which is gapped and exhibits alternating single- and double-valence bond singlets, and a fully dimerized phase, where three valence bond singlets form on every alternate bond \cite{chepiga2020floating,rachel2009spin}. The transitions between these phases are of various types - KT between the critical and partially dimerized phases, first order between the partially and fully dimerized phases, and either a first- or second-order between the critical and fully dimerized phases, depending on the parameters \cite{chepiga2020floating}.

A notable feature of frustrated spin chains is the presence of fractionalized excitations (spinons) in the spectra \cite{bethe1931theorie, clozieaux1962spin, muller1981quantum, bougourzi1998exact, lake2005quantum, des1962spin,rachel2009spin}, particularly in the vicinity of first-order transition points \cite{vanderstraeten2020spinon,sharma2025bound}. A paradigmatic example is provided by the spin-\(1/2\) \(J_1\)-\(J_2\) chain, where spinon domain-wall excitations appear at the MG point~\cite{caspers1984majumdar,ferrari2018dynamical,white1996dimerization}, with spinon bound states emerging near \(k = \pi/2\) \cite{lavarelo2014spinon,shastry1981excitation}. Similar domain-wall continua are expected in spin-1 and spin-\(3/2\) \(J_1\)-\(J_3\) chains near their respective MG points, where the ground state is fully dimerized and exactly solvable. By analogy with the spin-\(1/2\) chain, these domain-wall excitations are anticipated to be deconfined, potentially exhibiting bound states.

Another particularly interesting class of fractional excitations is that of the spinons that appear as domain walls between distinct valence bond solid (VBS) configurations and become deconfined at first-order transitions. Away from these transitions, these spinons experience confinement, binding into composite excitations that transform the excitation spectrum from a spinon continuum into a series of discrete bound-state modes \cite{vanderstraeten2020spinon}. Explicit signatures of fractionalization, including spinon deconfinement, have been observed in the spin-1 \(J_1\)-\(J_2\) chain at the first-order quantum phase transition near \(J_2 \approx 0.76 J_1\) \cite{kolezhuk1996first, pixley2014frustration, chepiga2016dimerization}, which separates distinct VBS phases~\cite{sharma2025bound}. Recent studies of the spin-\(3/2\) \(J_1\)-\(J_2\) chain further highlight this behavior~\cite{sharma3by2}, revealing that the excitation spectrum in the partially dimerized phase is dominated by deconfined spinons, with magnon-like resonances appearing within the spinon continuum. In \(J_1\)-\(J_2\)-\(J_3\) spin-1 and spin-\(3/2\) chains, first-order transitions, between the Haldane and dimerized phases for spin-1, and between partially and fully dimerized phases for spin-\(3/2\), are also expected to exhibit analogous confinement and fractionalization dynamics.

In this paper, we study the dynamical properties and provide a deeper understanding of the elementary excitations in the $J_1$-$J_2$-$J_3$ spin chains using state-of-the-art time-dependent density matrix renormalization group (tDMRG) techniques. In particular, we compute the dynamical structure factor (DSF) across various phases and phase transitions for spin-$1$ and spin-$3/2$ chains. Our goal is to understand the low-energy excitations in terms of magnons, spinons, and domain walls that arise in VBS configurations. To this end, we use the single mode approximation (SMA) to compute quasiparticle dispersions and compare them to the high-intensity features in the DSF. Our primary objectives are to:
\begin{enumerate}
    \item Explore the DSF of the $J_1$-$J_3$ model in the fully dimerized phase of the spin-$1/2$, spin-$1$, and spin-$3/2$ chains around their respective MG points.
    \item Characterize the excitations in the DSF along the transition line from the Haldane phase to the dimerized phase in the spin-1 chain as the phase transition changes its nature from first order to second order, resulting in closing of the spectral gap.
    \item Demonstrate the confinement of spinons across first-order phase transitions in both spin-1 and spin-3/2 chains.
\end{enumerate}

The remainder of this paper is structured as follows. In Sec. \ref{Methods}, we describe the numerical methods used to obtain the DSF, focusing on the tDMRG and the SMA and describe the construction of excitations in the VBS states. In Sec.~\ref{sec:DSF}, we present the DSF of the $J_1$–$J_3$ model across spin-$1/2$, spin-$1$, and spin-$3/2$ chains, and discuss the evolution of the continua of domain-wall excitations and magnons as a function of the magnitude of spin. In Sec.~\ref{spin1_J1_J2_J3}, we focus on the spin-$1$ chain and explore the evolution of DSF spectra along the phase transition between the Haldane phase and the fully dimerized phase, highlighting signatures of criticality and incommensurability. In Sec.~\ref{spinon_confinement}, we present the DSF for the first-order transitions in both spin-$1$ and spin-$3/2$ systems, where we uncover compelling evidence of spinon confinement as the spinon continua transform into discrete bound states. We conclude in Sec.~\ref{summary} with a summary of our findings.

\section{Methods}
\label{Methods}
\subsection{Time-dependent DMRG}
\label{tDMRG_benchmarking}

The DSF is very effective for probing the low-energy excitation spectrum of quantum spin systems. It serves as a direct link between theory and inelastic neutron scattering (INS) experiments. The definition of the DSF is
\begin{equation*}
S^{\alpha, \alpha}(k,\omega) = \int dt \, e^{-i \omega t} \sum_{r_i,r_j} e^{i k(r_i-r_j)} \bra{\psi_0} S^\alpha_{r_i} (t) S^\alpha_{r_j} \ket{\psi_0}, 
\label{eq:dsf}
\end{equation*}
where $\alpha = x, y, z$, $r_i$ and $ r_j$ are the site indices of the spin chain, and $\ket{\psi_0}$ is the ground state of the system. Due to the absence of anisotropy in the $J_1 - J_2- J_3$ Heisenberg spin chain, we concentrate only on the longitudinal component $S^{zz}(k,\omega)$ of the DSF, as the other components will be the same. To extract the DSF numerically, we follow a two-step procedure: (a) obtaining a high-precision ground state wavefunction, and (b) computing the real-time evolution of the state following a local spin perturbation to the ground state.  

The DMRG algorithm \cite{white1992density, white1993density, schollwock2011density} has been very effective in obtaining accurate ground states of strongly correlated one-dimensional systems. For this study, we perform iterative sweeps using a two-site DMRG algorithm. We report the system sizes and bond dimensions used, and quantify convergence through the variance of the ground state energy per site. For the spin-$1/2$ \( J_1 \)–\( J_2 \) chain, simulations were performed with a system size of 300 and bond dimension of 350, achieving energy variance between \(10^{-11}\) and \(10^{-13}\). At \( J_2 = 0.5 \), a smaller bond dimension of 100 was already sufficient to yield strong convergence. While low entanglement near exactly dimerized points allows smaller bond dimensions, a larger one is still needed to capture entanglement growth during time evolution. In the spin-$1$ \( J_1 \)–\( J_3 \) chain, we used a system size of 120 with bond dimension 120, obtaining ground-state energy variances per site in the range \(10^{-7}\) to \(10^{-10}\). In particular, at the MG point \( J_3 = J_1/6 \), the variance is the lowest with \(10^{-12}\). For the spin-$3/2$ \( J_1 \)–\( J_3 \) chain, we used system size of 150 sites and bond dimensions between 170 and 220, achieving variance per site ranging from \(10^{-11}\) to \(10^{-6}\). For higher values of  $J_3$, i.e. deep in the fully dimerized phase, the ground state is determined with a variance per site of \(10^{-11}\). Along the line of first-order transitions of the spin-1 \( J_1 \)–\( J_2 \)–\( J_3 \) chain, calculations were performed for system size 120 with bond dimensions of 120–150, yielding variances in the range \(10^{-5}\) to \(10^{-8}\). At fixed \( J_2 = 0.4 \), we used a bond dimension of 150 and achieved variance per site between \(10^{-7}\) and \(10^{-9}\). Finally, for the spin-3/2 \( J_1 \)–\( J_2 \)–\( J_3 \) chain, simulations at system size 150 and bond dimension 220 resulted in variance per site from \(10^{-4}\) to \(10^{-9}\), with improved precision down to \(10^{-11}\) for large \( J_3 \). To ensure reliable ground-state convergence, between 35 and 45 DMRG sweeps were performed in each case.

After the ground state is determined, a local spin operator is applied and then the real-time spin dynamics of the system is simulated using the tDMRG, also called time-evolving block decimation (TEBD)~\cite{white2004real,white2008spectral, white1992density, white1993density, schollwock2005density, schollwock2011density}. The full time evolution of a quantum state is governed by the unitary operator
\begin{eqnarray}
U(t) = {\left\lbrack U(\delta t)\right\rbrack}^{n},~\text{with}~U(\delta t) = e^{-iH\delta t},
\end{eqnarray}
 where $U(\delta t)$ is the time evolution for short times and the final time $t = n\delta t$. The $U(\delta t)$ is approximated using the second order Trotter–Suzuki decomposition~\cite{vidal2004efficient, daley2004time, feiguin2005time}. This scheme strikes a balance between accuracy and computational efficiency and is well suited to studying real-time quantum dynamics in one-dimensional systems. The choice of $\delta t$ is crucial for balancing accuracy and efficiency. In our simulations, a time step of \( \delta t = 0.02/J_1 \) was used for the spin-1/2 and spin-1 chains. For the spin-3/2 chains, a slightly larger time step of \( \delta t = 0.05/J_1 \) was used, optimized for efficient evolution while keeping the numerical errors low.

The entanglement entropy of the time-evolved state increases approximately linearly with time. This leads to exponential growth of computational cost, requiring systematic truncation of the bond dimension to prevent memory overflow. We address this by fixing the bond dimension across all bonds during the time evolution. The truncation error incurred by dropping the smaller singular values is smaller than the Trotter error incurred in the time evolution.

Applying a local spin operator to the ground state excites a superposition of eigenstates with finite energy and momentum. This perturbation spreads across the chain in a light-cone-like manner, generating a wavefront of correlations. The speed at which these wavefronts travel depends on the system's coupling ratios and the spin of system. In order to avoid the numerical artifacts due to the boundaries, we ensure that the system is evolved only until the wavefronts reach the edges of the finite chain. In some cases where the speed of the wavefronts was very slow, the time-evolution was stopped before the wavefronts reach the edges. For the $J_1$–$J_2$ spin-$1/2$ chain, we evolved up to times between $70/J_1$ and $150/J_1$. For the $J_1$–$J_3$ spin-1 chain, the evolution time ranged from $30/J_1$ to $90/J_1$. At the MG point, where the velocity is particularly low, we did the time evolution up to $150/J_1$. For the $J_1$–$J_3$ spin-$3/2$ chain, we used evolution times between $40/J_1$ and $75/J_1$. Along the line of first-order transitions of the spin-1 $J_1$–$J_2$–$J_3$ chain, we evolved to $50/J_1$–$100/J_1$. For the spin-1 $J_1$–$J_2$–$J_3$ chain across the first-order transition at $J_2 = 0.4 J_1$, the evolution time was $50/J_1$–$130/J_1$. Finally, for the spin-$3/2$ $J_1$–$J_2$–$J_3$ chain across the first-order transition at $J_2 = 0.328 J_1$, evolution times between $50/J_1$ and $125/J_1$ were used.\par

To compute the DSF, we perform a double Fourier transform in both space and time. However, since time evolution is limited to a finite window, the Fourier transformation introduces ringing artifacts because of sharp cutoffs. To mitigate this, we employ a Gaussian filter for the time-dependent correlation function,
\begin{equation}
G(t) \propto e^{-\frac{t^2}{2\sigma^2}}.
\end{equation}
This reduces artificial oscillations at the cost of introducing spectral broadening. After testing different values, we set the filter parameter to $\sigma = 0.276 t_f$, ensuring an optimal balance between resolution and suppression of finite-time artifacts.

The combination of DMRG for ground state preparation and tDMRG for real-time evolution provides a highly effective framework for extracting spectral functions for the $J_1$–$J_2$–$J_3$ spin chains. The extracted spectra will be discussed in the subsequent sections, where we analyze the interplay between fractionalization, confinement, and magnon dynamics.

\subsection{Valence Bond Solid Ansatz}

\begin{figure}[htbp] 
    \centering
    \includegraphics[width=\columnwidth]{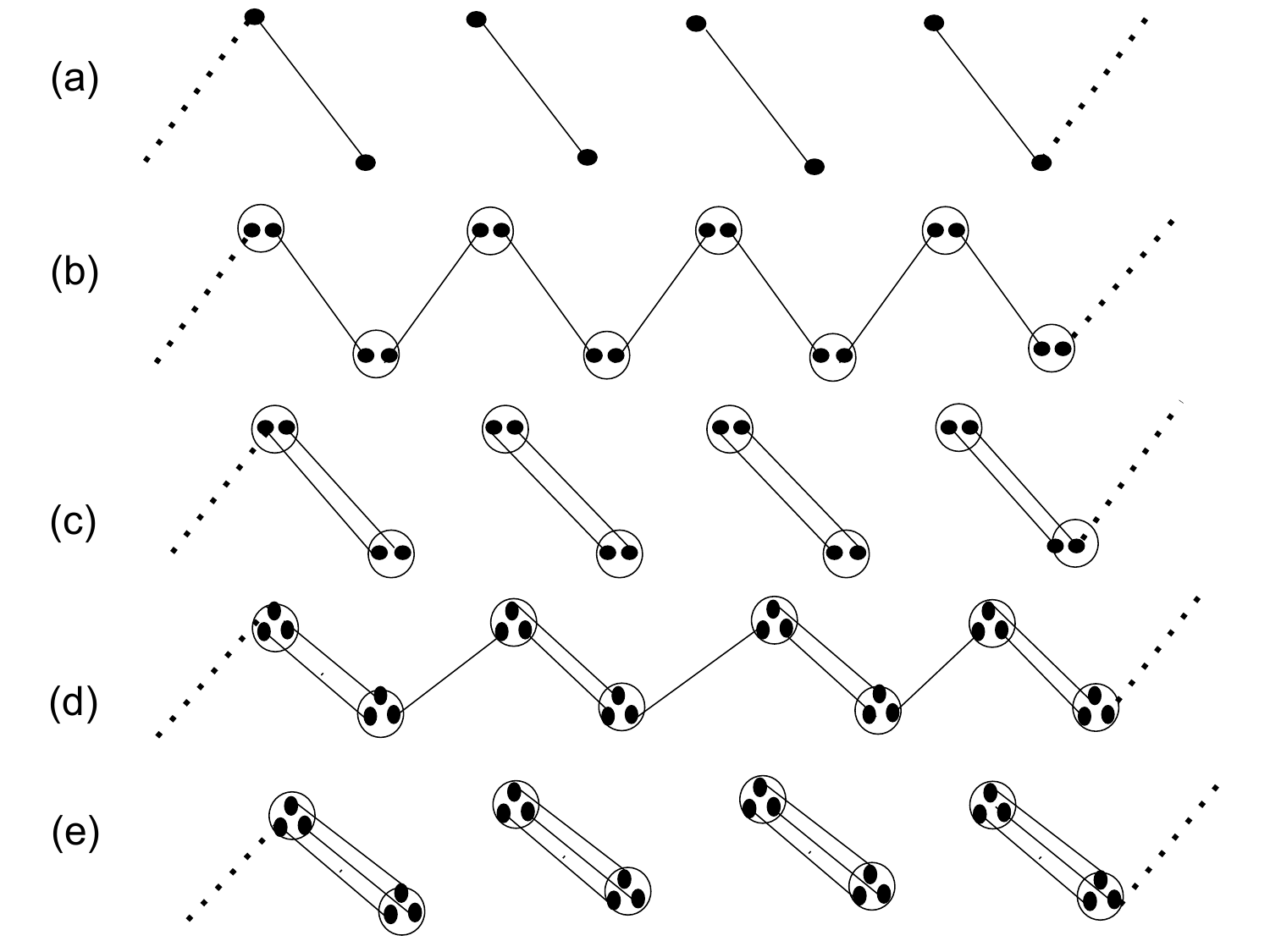} 
     \caption{VBS states for different spin chains. 
    (a) Spin-1/2 chain in a dimerized state with alternating singlet bonds. 
(b) Spin-1 chain in the Haldane state, where each valence bond comprises a single singlet. 
(c) Spin-1 chain in the dimerized state with two singlet bonds per valence bond. 
(d) Spin-3/2 chain in a partially dimerized state with a mixture of one and two singlets per valence bond. 
(e) Spin-3/2 chain in a fully dimerized state, with three singlet bonds per valence bond. At each site, the \(2S\) spin-\(1/2\) constituents are symmetrically combined to form a total spin \(S\). The chains are drawn in a staggered layout purely for visual clarity, to distinguish even and odd sites}
    \label{fig:vbsgs}
\end{figure}

\begin{figure}[htbp] 
    \centering
    \includegraphics[width=0.93\columnwidth]{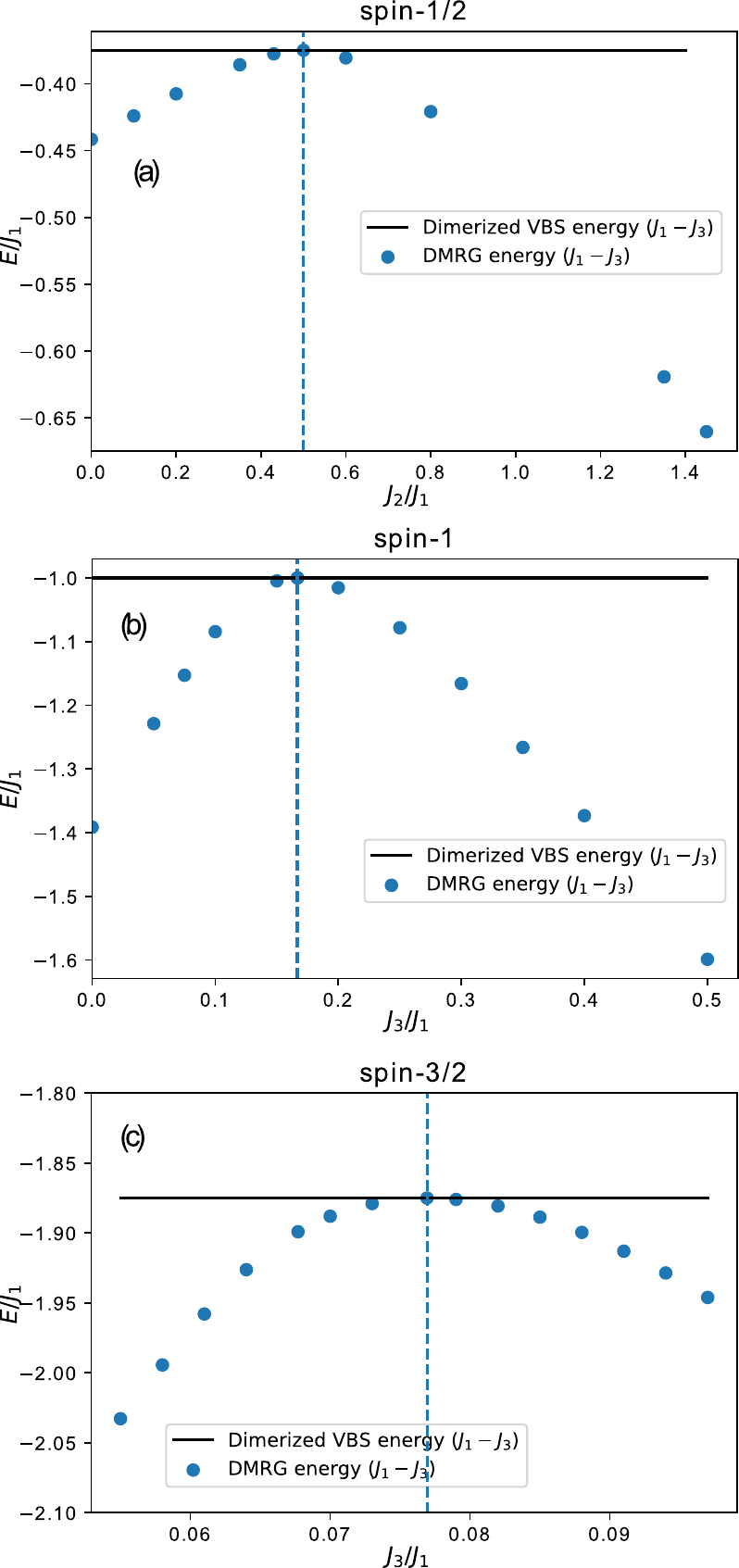} 
    \caption{Comparison of the dimerized VBS energy expectation values (\(E/J_1\)) with the ground state energy from DMRG for the (a) spin-1/2 \(J_1\text{--}J_2\) chain, (b) spin-1 \(J_1\text{--}J_3\) chain, and (c) spin-3/2 \(J_1\text{--}J_3\) chain. The close match near the exact dimerization points ($J_3 = J_1/6$ for spin-1 and $J_3 = J_1/13$ for spin-3/2) underscores the accuracy of the VBS ansatz in these regimes. DMRG energies were obtained on systems of 120 sites for spin-$1$ chain and 150 sites for spin-$3/2$ chain.}
    \label{fig:ene}
\end{figure}

\begin{figure}[htbp] 
    \centering
    \includegraphics[width=\columnwidth]{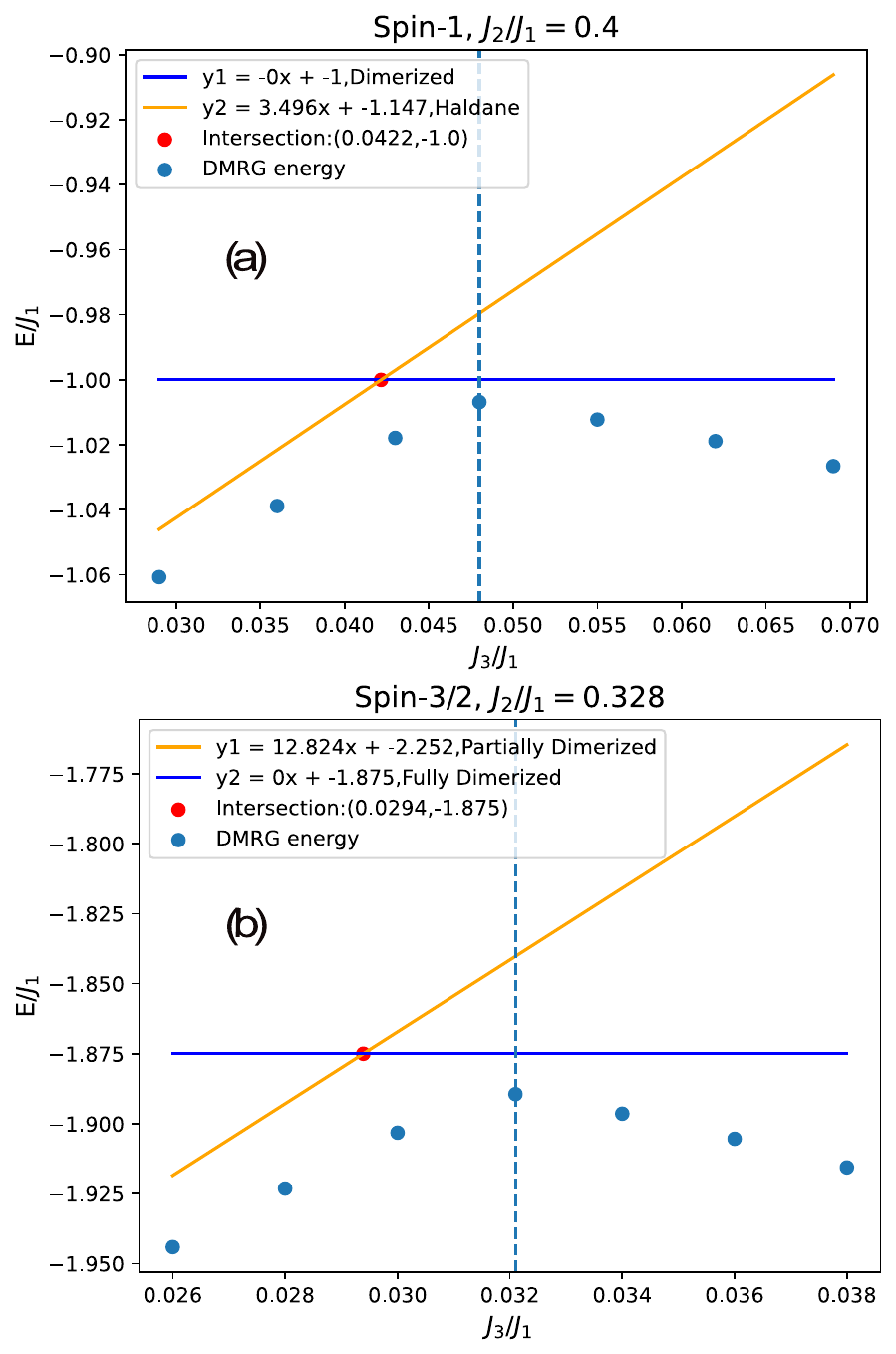} 
    \caption{(a) Comparison of VBS energies and DMRG ground-state energies (\(E/J_1\)) for the spin-1 \(J_1\)–\(J_2\)–\(J_3\) chain at fixed \(J_2 = 0.4J_1\) (cut C1, see Fig.~\ref{phasediags}a), using dimerized and Haldane VBS wavefunctions. (b) Same for the spin-3/2 chain at \(J_2 = 0.328J_1\) (cut C2, see Fig.~\ref{phasediags}b), using fully and partially dimerized VBS states. In both cases, the crossing point of the two VBS energy lines closely approximates the DMRG-determined first-order transition point extracted from the phase diagrams in Fig.~\ref{phasediags}, indicated in the plots above by vertical dashed lines~\cite{chepiga2016dimerization,chepiga2020floating}. DMRG energies were obtained on systems of 120 sites for spin-$1$ chain and 150 sites for spin-$3/2$ chain.}
    \label{fig:enej1j2j3}
\end{figure}

The VBS \emph{ansatz} offers an elegant and intuitive description of gapped ground states in one-dimensional quantum spin chains \cite{affleck1989quantum,kolezhuk1996first,rachel2009spin}. These wavefunctions can be constructed by hosting 2S spin-1/2 degrees of freedom at each site of the spin chain and forming singlet bonds between spin-1/2s on neighboring sites. Then, the 2S spin-1/2 degrees of freedom are symmetrically combined to form a spin-S degree of freedom at each site of the spin chain. Here, “symmetrically combined” refers to the projection of the $2S$ spin-$1/2$ constituents onto the fully symmetric spin-$S$ irrep. Singlets are formed between neighboring spins in a bond pattern appropriate to the regime, such as Haldane, fully dimerized, or partially dimerized. Depending on the spin value and the interaction parameters, different bond patterns arise. This construction is non-variational (i.e., it contains no free parameters) and is used as a controlled starting point to generate approximate excitations via the SMA.

Figure~\ref{fig:vbsgs} illustrates the representative VBS states across various spin magnitudes. For the spin-1/2 chain, the dimerized state consists of alternating singlet bonds on adjacent links [Fig.~\ref{fig:vbsgs}(a)]. For spin-1 chains, two prominent VBS configurations arise: the Haldane state [Fig.~\ref{fig:vbsgs}(b)], where each spin is connected to its neighbors via a single singlet, and a dimerized phase [Fig.~\ref{fig:vbsgs}(c)], characterized by two singlets per bond. The spin-3/2 chain exhibits an even richer structure, allowing for both partially dimerized [Fig.~\ref{fig:vbsgs}(d)] and fully dimerized [Fig.~\ref{fig:vbsgs}(e)] phases. In the partially dimerized state, there is a bond-alternating pattern of two and one singlets. In contrast, the fully dimerized state places all three singlet bonds on every alternate bond (e.g., all on even bonds), leaving the intervening bonds uncoupled. This results in a maximally dimerized state. Both configurations break translational symmetry and are doubly degenerate.

To quantitatively assess the validity of the VBS wavefunctions, we compute the energy per site (\(E/J_1\)) by evaluating the expectation value of the relevant system Hamiltonian using the VBS ansatz and dividing by the total number of lattice sites. The ground-state energy from DMRG is computed in the same way by taking the expectation value of the Hamiltonian with the numerically obtained ground state and normalizing by the system size. These energies, referred to as ``VBS energy'' and DMRG energy respectively, are compared in Figs.~\ref{fig:ene} and~\ref{fig:enej1j2j3}. The spin-1/2 chain provides a useful benchmark, where the energy expectation value of the fully dimerized VBS state exactly matches the ground state energy at the MG point, $J_2 = 0.5 J_1$ [Fig.~\ref{fig:ene}a].

For the spin-1 $J_1$–$J_3$ Heisenberg chain, the comparison shown in Fig.~\ref{fig:ene}b reveals excellent agreement between the energies of the fully dimerized VBS wavefunction and the DMRG ground state near the exactly dimerized point or the MG point $J_3 = J_1/6$, as expected. This validates the ansatz [Fig.~\ref{fig:vbsgs}c] in this regime. In Fig.~\ref{fig:ene}c, a similar comparison of the energies is performed for the spin-3/2 $J_1$–$J_3$ chain. Again, close agreement is observed, consistent with expectation near the MG point $J_3 = J_1/13$.

For the spin-1 $J_1$–$J_2$–$J_3$ chain at fixed $J_2 = 0.4J_1$ (cut C1, see Fig.~\ref{phasediags}a), we compute the energy expectation values for both the dimerized and Haldane VBS states. As shown in Fig.~\ref{fig:enej1j2j3}a, the crossing point between the two energy curves aligns closely with the first-order transition point obtained from DMRG \cite{chepiga2016dimerization}, providing strong support for the applicability of VBS wavefunctions in noting phase transitions.

An analogous situation occurs in the spin-$3/2$ $J_1$–$J_2$–$J_3$ chain, where the competition is between the partially dimerized and fully dimerized VBS phases. Figure~\ref{fig:enej1j2j3}b shows the corresponding energy comparison at fixed $J_2 = 0.328J_1$ (cut C2, see Fig.~\ref{phasediags}b). The crossing of the energy curves marks the location of a first-order transition, again consistent with the DMRG calculations \cite{chepiga2020floating}. This highlights how the VBS approach captures not only ground state energies but also the competing phases of the model.

In summary, the VBS ansatz provides a physically motivated and qualitative description of the ground states in frustrated spin chains across a range of spin values. Its success in capturing the essential features of the ground state, as reflected in its close agreement with DMRG ground state energies in specific parameter regimes, and characterizing symmetry-broken phases points to its usefulness as a framework for analyzing the excitations in such phases. In the following sections, we construct and analyze a rich variety of excitations, including magnons, domain walls, and spinons, using the SMA on representative VBS states.

\subsection{SMA and excitations in VBS States}
\label{SMAsection}

\begin{figure}[htbp] 
    \centering
    \includegraphics[width=\columnwidth]{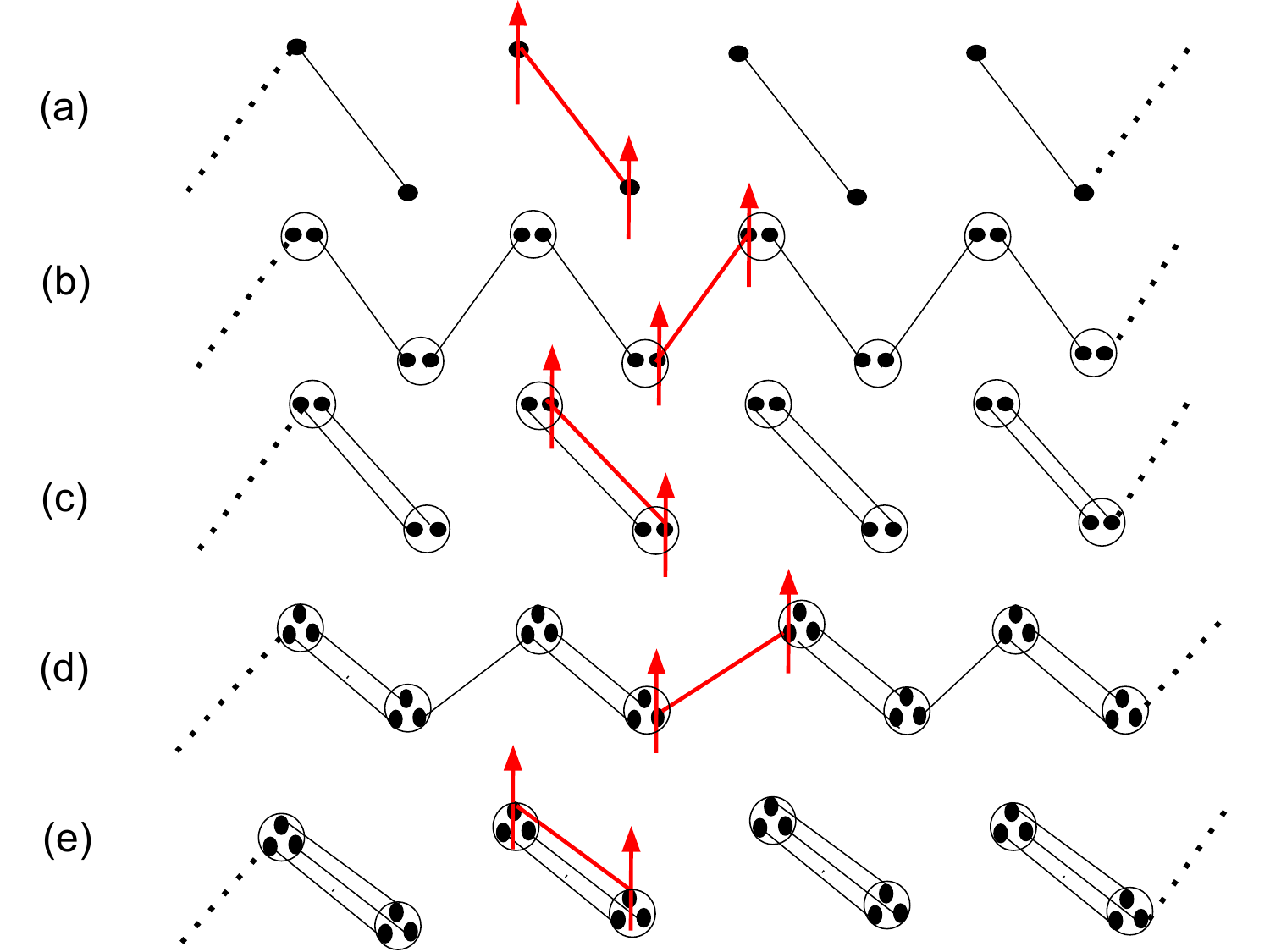} 
       \caption{Magnon excitation created by flipping a singlet bond in VBS states. 
Figs. (a)–(e) show representative magnon excitations in various spin chain VBS configurations: 
(a) spin-1/2 dimerized state, 
(b) spin-1 Haldane state, 
(c) spin-1 dimerized state, 
(d) spin-3/2 partially-dimerized state, and 
(e) spin-3/2 fully dimerized state. 
In each case, a singlet bond is promoted to a triplet, generating a local $S=1$ excitation that can propagate as a magnon mode.
}
    \label{fig:mag}
\end{figure}

\begin{figure}[htbp] 
    \centering
    \includegraphics[width=\columnwidth]{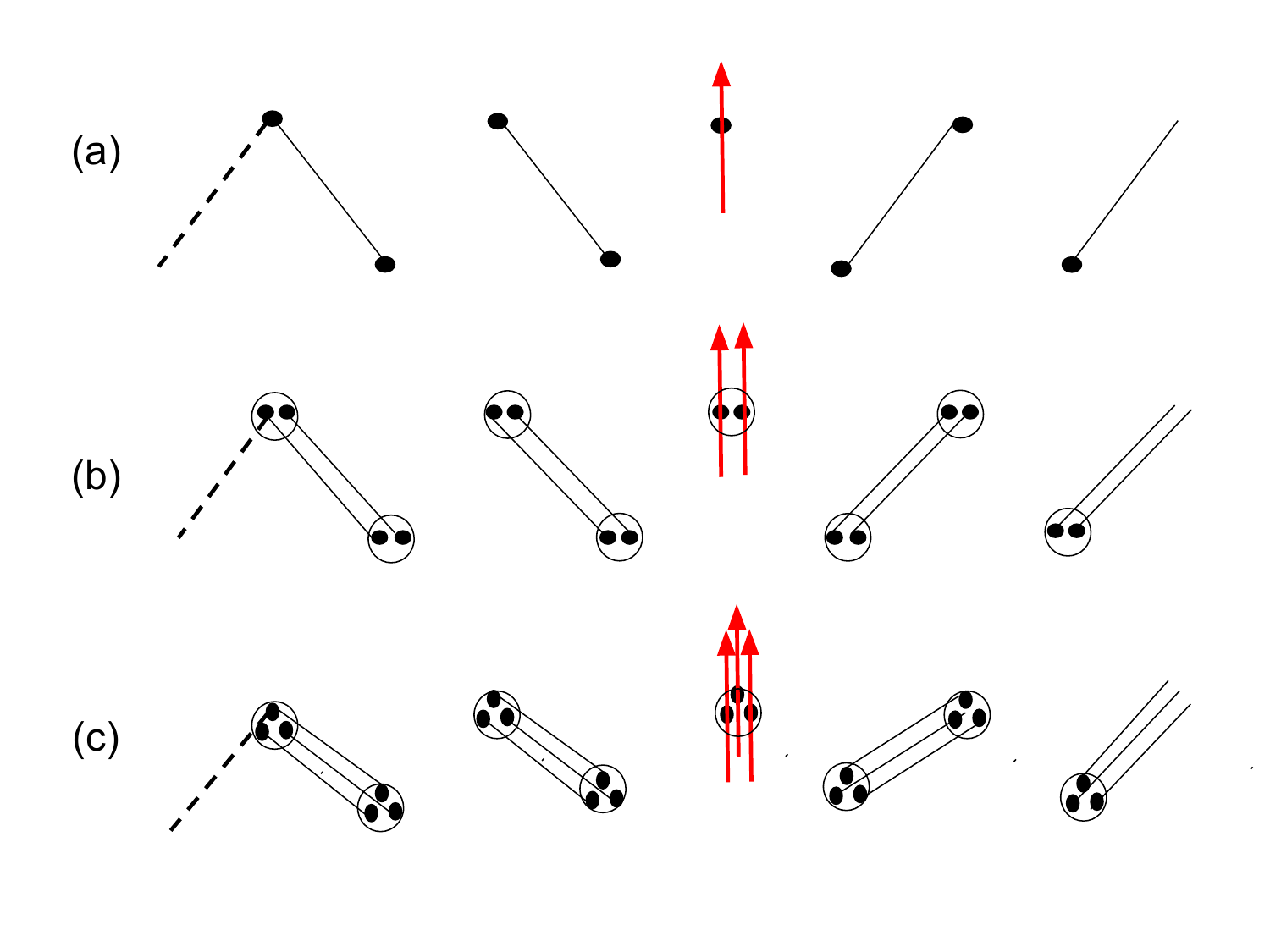} 
    \caption{Domain-wall excitations within dimerized VBS states that are discussed in the main text. (a)–(c) show domain walls for spin-1/2, spin-1, and spin-3/2 chains, where a mismatch between two degenerate dimerized states creates a solitonic defect. The red arrows are not meant to represent a specific $S^z$ component, but rather the total spin-$S$ degree of freedom.}
    \label{fig:domainwall}
\end{figure}

\begin{figure}[htbp] 
    \centering
    \includegraphics[width=\columnwidth]{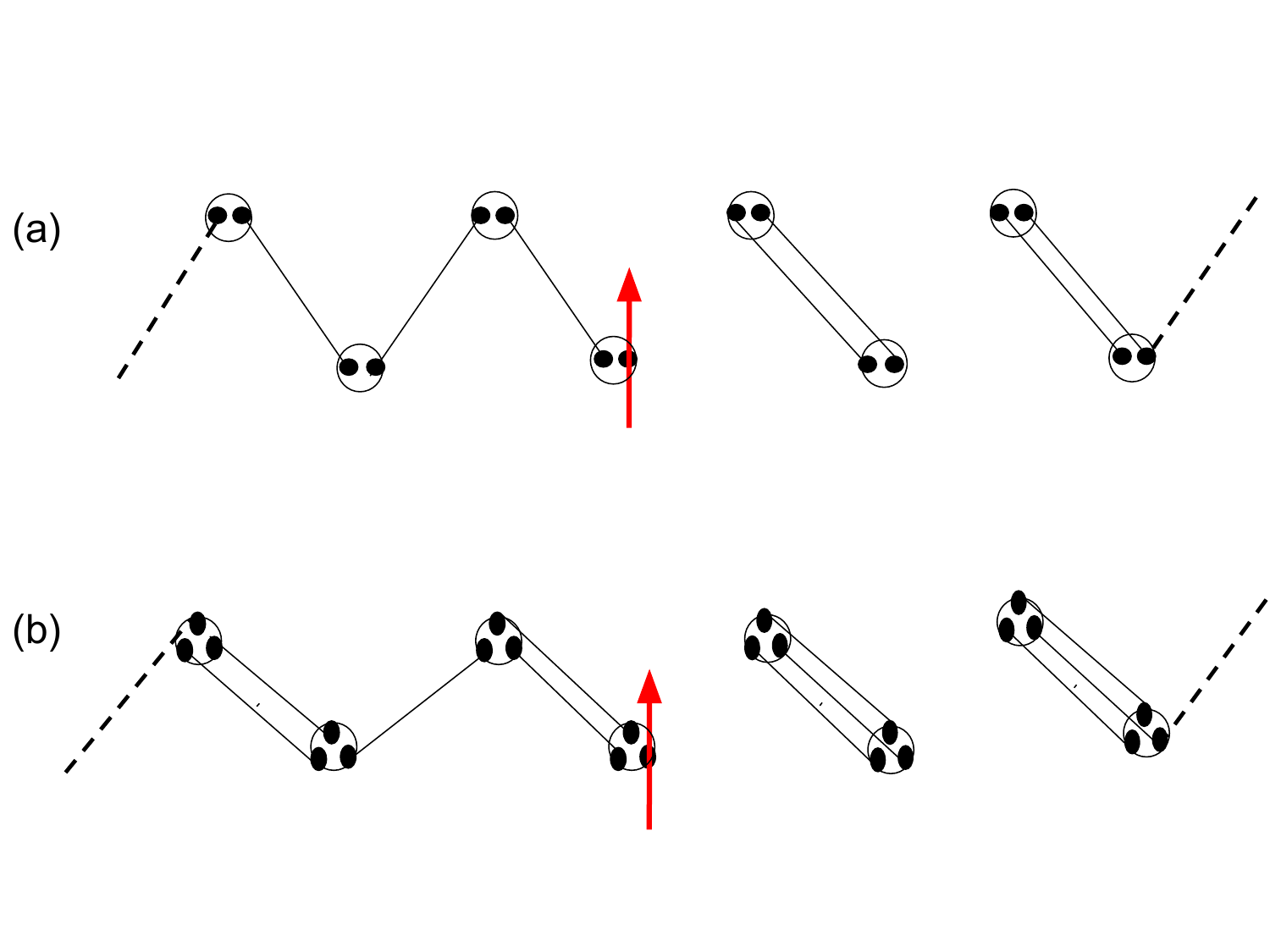} 
        \caption{Spinon domain wall between two distinct VBS states. 
(a) Domain wall at the interface between a Haldane VBS state and a dimerized state in the spin-1 chain. 
(b) Domain wall between a partially dimerized and a fully dimerized VBS state in the spin-3/2 chain. 
}
    \label{fig:domainwalldis}
\end{figure}

The SMA \cite{bijl1940lowest, feynman1954atomic, girvin1986magneto} provides a framework to analyze the dispersion of low-energy excitations by constructing momentum resolved states from local spin operators acting on the ground state  \cite{auerbach1998interacting}. When a momentum-resolved spin operator is applied to the ground state, it produces a state that approximates an excited eigenstate. The expectation value of the Hamiltonian with respect to this state, divided by its norm, provides an estimate of the corresponding excitation energy \cite{auerbach1998interacting}. In many systems, this approach captures the essential features of elementary excitations and has been particularly successful in describing magnon dispersions in VBS systems, such as the Affleck-Kennedy-Lieb-Tasaki (AKLT) model~\cite{arovas1988extended, auerbach1998interacting, fath1993solitonic}. 

For a chain of length \( N \), an SMA state carrying momentum \( k \) is constructed as:
\begin{eqnarray}
\ket{k} = \frac{1}{\sqrt{N}}\sum\limits_j e^{i k r_j}\Omega_j|\psi_0\rangle,
\label{Eq_sma_state}
\end{eqnarray}
where \( \Omega_j \) is a local operator acting on site \( j \), and \( \ket{\psi_0} \) is the ground state. The choice of \( \Omega_j \) depends on the nature of the excitation being studied. For instance, to model a magnon excitation, one typically takes \( \Omega_j = S_j^z \), which promotes a singlet bond to a triplet. Alternatively, one can use \( \Omega_j = S_j^{+} \) or \( S_j^{-} \), which generate transverse spin excitations by raising or lowering the total \( S^z \) quantum number by one unit. The corresponding dispersion relation is obtained from the relation:
 \begin{eqnarray}
\omega (k)= \frac{\bra{k}H\ket{k}}{\langle k\ket{k}}-E_0,
\label{Eq_sma_defn}
\end{eqnarray}
with $E_0$ being the ground state energy. Upon substitution of Eq. \ref{Eq_sma_state} into Eq. \ref{Eq_sma_defn} and simplification, the resulting expression can be obtained in the form:
\begin{eqnarray}
\omega(k)=\frac{a_0+\sum\limits_{n=1}^{N/2}a_n\cos(nk)}{1+\sum\limits_{n=1}^{N/2}b_n\cos(nk)} - E_0
\label{Eq_sma_defn_trig}
\end{eqnarray}
The terms \( a_n \) reflect contributions from Hamiltonian matrix elements between states with the quasiparticle displaced by \( n \) sites, while \( b_n \) quantify the nonorthogonality between such states. 

Using this framework, we examine three distinct classes of  quasiparticle excitations: magnons, domain walls within a dimerized VBS state, and domain walls between different competing VBS states.

\begin{enumerate}
\item {\it Magnon Excitation}:

Magnon excitations in a VBS state can be constructed by locally promoting a singlet bond into a triplet. In Fig.~\ref{fig:mag} examples of such magnon states are shown for different spin chains. Such a process introduces a local $S=1$ excitation that can propagate across the chain, forming a dispersive mode. A particularly insightful way to generate a magnon is to apply the Fourier-transformed spin operator $S^z_{k}$ on the VBS state \cite{kolezhuk1997variational, fath1993solitonic,arovas1988extended}:
\[
|\psi_k^{\text{mag}}\rangle = S^z_{k} |\psi_0\rangle,
\]
where $|\psi_0\rangle$ denotes the VBS state (Haldane or dimerized, depending on $J_3$). This corresponds to choosing \( \Omega_j = S^z_j \) in Eq.~\ref{Eq_sma_state}.

Following Refs.~\cite{kolezhuk1997variational, fath1993solitonic}, we note that acting with $S^z_{k}$ on a VBS state creates a coherent linear combination of local triplet excitations, effectively forming a one-magnon wave packet. Alternatively, in real space, we can construct a magnon by explicitly flipping a singlet bond to a triplet and then taking a Fourier transform to result in a state with a definite momentum. In this context, the operator \( \Omega_j \) in Eq.~\ref{Eq_sma_state} represents a local excitation that converts the singlet bond between sites \( j \) and \( j+1 \) into a triplet in VBS state $|\psi_0\rangle$. The energy of such a state is then computed using Eq.~\ref{Eq_sma_defn}, where \( E_0 \) denotes the expectation value of the Hamiltonian in the VBS state. This procedure provides a magnon dispersion relation, which can be directly compared with the low-energy features of the DSF.

In the spin-3/2 $J_1$–$J_2$–$J_3$ chain, the ground state structure again varies with parameters. To study magnon excitations in the partially dimerized phase, we construct a real-space excitation by converting one of the singlet bonds in the VBS configuration into a triplet. In the fully dimerized phase at larger $J_3$, a simpler alternative is to act directly with the $S^z_{k}$ operator on the VBS state. The physics for the magnon mode in higher spin chains remains essentially same as in spin-$1$ chain. In both spin-1 and spin-3/2 cases, these magnon modes can be used to approximate the lowest-lying excitations visible in the DSF.

\item \label{item:domain1} \textit{Domain-Wall Excitations in Dimerized Phases}:
In addition to conventional magnon modes, an equally significant class of excitations arises from domain walls, which naturally occur in dimerized spin chains due to the presence of degenerate ground states \cite{lavarelo2014spinon,rachel2009spin}. These excitations can be realized by making use of the two-fold degeneracy of the dimerized state. By translating a VBS configuration by one lattice site, one obtains a state orthogonal to the original (in the thermodynamic limit), with a mismatched singlet pattern. The junction between the two configurations is the domain wall. In real space, this corresponds to placing the singlet covering of one region in a pattern shifted by one site relative to the other, creating an unpaired spin at the interface. Figure~\ref{fig:domainwall} illustrates these domain walls across different spin chains. Furthermore, the domain wall can move throughout the chain leading to a dispersive excitation band.

The domain wall appears as a free spin situated at the boundary between the two dimer backgrounds. For spin-1 chains, this free spin carries an $S=1$ quantum number. For spin-3/2 chains, the dimerized state consists of three singlets per dimer, and a domain wall induces a local spin-$3/2$ excitation.

The domain walls cannot be generated by local spin operators such as $S^z$ alone. However, by explicitly preparing states with domain boundaries, either numerically or analytically using the VBS ansatz, one can study their propagation and energy cost. The dispersion relation of such domain walls can be obtained using the SMA. Specifically, one constructs the momentum-labelled state by taking the Fourier superpositions of real-space configurations where the domain wall is centered at different positions:
\[
|\psi_k^{\text{dw}}\rangle = \sum_{j} e^{ikj} |\text{DW}_j\rangle,
\]
where $|\text{DW}_j\rangle$ is the dimerized chain with a domain wall centered at site $j$. The excitation energy is then computed using Eq.~\ref{Eq_sma_defn}, by replacing \( |k\rangle \) with \( |\psi_k^{\text{dw}}\rangle \), and with \( E_0 \) taken as the expectation value of the Hamiltonian in the VBS state.

\item \label{item:domain2} \textit{Domain-Wall Excitation (between different VBS states)}:
Another class of elementary excitations emerges at the interface between two competing VBS states~\cite{sharma2025bound,vanderstraeten2020spinon,rachel2009spin}. They are shown in Fig.~\ref{fig:domainwalldis}. For the spin-1 chain, these two states are the Haldane state and the dimerized state, while for the spin-3/2 chain, they are the partially dimerized and fully dimerized states. The domain wall formed at the junction of these VBS patterns constitutes a local excitation that carries a fractional spin quantum number, $S=1/2$, and is thus identified as a spinon. The domain wall separating these two ground-state sectors can propagate through the system. In this context, the domain wall is not merely a defect, but a fractionalized quasiparticle. Similar to the quasiparticle excitations discussed earlier, these domain-wall configurations offer valuable insight into the low-energy properties of the system.

As for the excitation described in Item \ref{item:domain1}, constructing such an excitation will require the application of a nonlocal operator that converts one VBS pattern into another across a local region but a closed-form expression for such an operator is generally intractable. These domain-wall states can be generated numerically or analytically by preparing variational wavefunctions consisting of sharp interfaces between the two states. The domain wall becomes energetically favorable near the first-order transition point, where the energies of the two competing states become degenerate. Specifically, in the spin-$1$ chain, this excitation corresponds to the lowest-energy state at the Haldane phase to fully dimerized phase transition point, and its energy increases rapidly when moving away from this point into either phase.

These domain walls are deeply related to magnons. By placing two spinon domain walls adjacent to each other, one restores a triplet bond and therefore, effectively reconstructs the magnon excitation discussed earlier. This connection highlights the fractionalized nature of the excitations. They carry half the spin of a magnon ($S=1$), similar to how spinons emerge in spin-$1/2$ chains. In the spin-$3/2$ chain, the same mechanism applies at the junction between the partially and fully dimerized states. In this case, the spinon is also a domain wall of spin-$1/2$ between the VBS state with alternating one-two and VBS state with three singlet bonds (see Fig.\ref{fig:domainwalldis} b). The transition point between these phases is expected to host deconfined spinons, which become confined into bound states away from it. Thus, these excitations are expected to provide a sensitive probe of phase competition and fractionalization in higher-spin systems.

\end{enumerate}

Because domain walls discussed in Items~\ref{item:domain1} and~\ref{item:domain2} are created in pairs and can propagate independently, they give rise to a two-particle continuum in the spectral function. The total energy and momentum of such a pair are given by:
\[
K = (k_1 + k_2)~\mathrm{mod}~2\pi, \quad E_K = E_{k_1} + E_{k_2},
\]
where \( k_1 \) and \( k_2 \) are the momenta, and \( E_{k_1} \), \( E_{k_2} \) are the energies of the individual domain walls. In later sections, we employ this approach repeatedly to obtain the two-domain-wall continua, using domain-wall dispersions computed via the SMA.

In practice, to construct the VBS ansatz for the excited states discussed in this section, we employed the matrix product state (MPS) framework, following the methodology detailed in the appendix of Ref.~\cite{sharma2025bound}.

\section{DSF of the $J_1 - J_3$ Model}
\label{sec:DSF}

\begin{figure*}[htbp]
    \centering
    \includegraphics[width=1\textwidth]{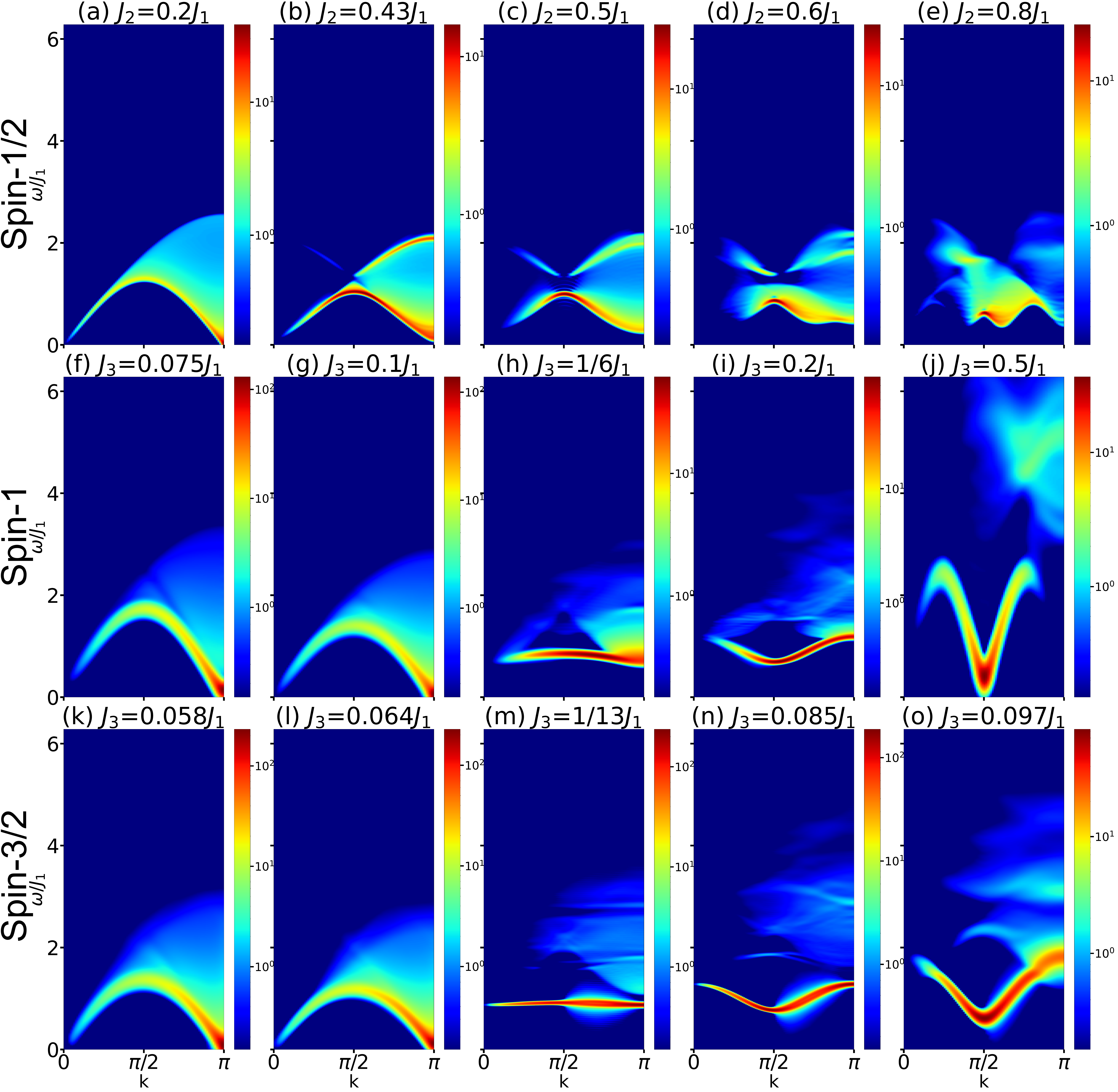} 
    \caption{(a–e) DSF of the $J_1$–$J_2$ spin-1/2 Heisenberg chain near the MG point. DSF of the $J_1$–$J_3$ Heisenberg chain with (f–j) spin-$1$ and (k–o) spin-$3/2$ across increasing values of $J_3$. As $J_3$ increases, both systems undergo phase transitions into fully dimerized states, at $J_3 = J_1/6$ for spin-1 and $J_3 = J_1/13$ for spin-3/2, marked by the domain-wall continua. Near the critical points ($J_3 \approx 0.111 J_1$ for spin-1 and $J_3 \approx 0.063 J_1$ for spin-3/2), the spectra reflect the deconfinement of spinon-like excitations\cite{affleck1987critical,michaud2012antiferromagnetic,michaud2013realization}. Nearly flat low-energy modes appear near the MG point. The simulations were performed on systems of 300 sites for spin-$1/2$ chain, 120 sites for spin-$1$ chain, and 150 sites for spin-$3/2$ chain. 
    }
    \label{fig:dsfsj1j3spin1and3by2}
\end{figure*}

In this section, we present and analyze the DSF spectra for the $J_1 - J_3$ Heisenberg model, comparing the excitation spectra for different spin values ($S=1/2, ~1, ~3/2$). The panels in Fig.~\ref{fig:dsfsj1j3spin1and3by2} illustrate the computed DSF for different values of $J_3$, with separate plots for spin-$1/2$, spin-1, and spin-3/2 chains. These spectra reveal key structural changes in the excitation continuum and provide crucial insights into the nature of excitations around the MG point of the spin chains.

To establish a reference, we first consider the spin-$1/2$ $J_1$–$J_2$ chain, which is equivalent to the $J_1$–$J_3$ model with $J_3 = 2J_2$. Figure~\ref{fig:dsfsj1j3spin1and3by2}(a–e) shows the DSFs. At $J_2/J_1 = 0.2$, the spectrum is gapless, consistent with the Tomonaga–Luttinger liquid phase, and follows the des Cloizeaux–Pearson form \cite{clozieaux1962spin}. For $J_2/J_1 \geq 0.43$, a gap opens and bound-state features begin to emerge. At the MG point ($J_2/J_1 = 0.5$), the ground state is exactly dimerized, and the DSF exhibits bound spinon excitations around $k = \pi/2$. For larger frustration, e.g., $J_2/J_1 = 0.6$, the spectral weight shifts away from $k = \pi$, in contrast to the smaller $J_2/J_1$ regime where it is concentrated at $k = \pi$. 
At and above $J_2/J_1 = 0.6$, non-monotonic behavior develops in the dispersion of the lower edge of the continuum, reflecting the emergence of incommensurate correlations. These features are consistent with the known behavior of the spin-$1/2$ $J_1$–$J_2$ chain, which undergoes a Kosterlitz–Thouless (KT) transition into a gapped dimerized phase at $J_2/J_1 \approx 0.2411$. The MG point corresponds to an exactly solvable limit with bound states centered at $k = \pi/2$, and for $J_2/J_1 > 0.52$~\cite{bursill1995numerical}, the system is known to exhibit incommensurate correlations.

In Fig.~\ref{fig:dsfsj1j3spin1and3by2}(f–j), the spin-1 DSFs exhibit a distinct evolution with increasing $J_3$. At small $J_3$ ($J_3 = 0.075 J_1$), the spectrum displays gapped excitations and a dispersive triplet magnon mode consistent with the Haldane phase. Although the spin gap is not easily discernible on the scale of the figure, it can be inferred from the fact that the dominant spectral weight at $k=\pi$ remains at finite energy and the slope of the lower spectral boundary decreases as $k \to \pi$. As $J_3$ increases, the Haldane gap closes near $J_3 \approx 0.1 J_1$ \cite{michaud2012antiferromagnetic}, and the spectrum approaches the des Cloizeaux–Pearson form, indicating a continuous transition. The spectral weight redistributes broadly, reflecting deconfined spinon behavior. Significant spectral changes occur between $J_3 = 0.1 J_1$ and $J_3 = 0.1666 J_1$, signaling the approach to a different phase. At the MG point ($J_3 = J_1/6$), the ground state is exactly dimerized, and the DSF shows a low-energy mode that is nearly dispersionless. The mode exhibits weaker dispersion compared to the magnon observed in the spin-\( 1/2 \) chain at \( k = \pi/2 \) at the MG point. Interestingly, remnants of the excitation continuum that appear near the critical point persist even at the MG point, particularly around \( k = \pi \) and energy \( \omega/J_1 \approx 1 \). The spectral features discussed above are consistent with the established phase diagram of the spin-1 $J_1$–$J_3$ chain, which includes the Haldane phase at small $J_3$ with a finite spin gap, a continuous transition into a dimerized phase at $J_3 \approx 0.111 J_1$ belonging to the SU(2)$_2$ WZW universality class with central charge $c=3/2$~\cite{michaud2012antiferromagnetic}. The generalized MG point at $J_3/J_1=1/6$, similar to its spin-\(1/2\) counterpart, is a disorder point beyond which the dominant wave vector deviates from $q=\pi$, and incommensurate short-range correlations develop. However, unlike the spin-\(1/2\) case, the incommensurate region in the spin-\(1\) $J_1-J_3$ chain is very narrow, and already at $J_3/J_1\approx0.185$ the dominant wave-vector reaches its new commensurate value $q=\pi/2$ \cite{chepiga2016spontaneous}. In Appendix~\ref{appendix:dsf_comm}, we present additional data and demonstrate that the switching from $\pi$ to $\pi/2$ in the DSF is not associated with a progressive development of incommensurability but happens abruptly, signaling a crossover between correlations at two different wave vectors.

The DSF for spin-3/2 chains is shown in Fig.~\ref{fig:dsfsj1j3spin1and3by2}(k–o). Here, the system remains gapless up to $J_3 \approx 0.064 J_1$, beyond which it becomes gapped. At and before $J_3 \approx 0.064 J_1$ the DSF displays a des-Cloizeaux Pearson like continuum, and beyond this point, sharp magnon modes are prominent feature in the DSF. The MG point for the spin-3/2 chain is located at $J_3 = J_1/13$. At this point, the spectrum shows nearly dispersionless mode similar to that seen in the spin-1 case. This mode displays even weaker dispersion compared to its counterpart in the spin-\( 1/2 \) chain at \( k = \pi/2 \), and also relative to the spin-1 case across the Brillouin zone at the MG point. In addition, in the DSF spectrum at the MG point the remeniscents of the continuuum which appeared at and before the critical point can still be seen near \( k = \pi \) and around the energy $\omega/J_1\approx1$. The observations discussed above are consistent with the phase structure of the spin-\(3/2\) \(J_1\)–\(J_3\) chain which is well established in the literature. The system remains in a gapless critical phase up to a continuous transition at \( J_3 \approx 0.063 J_1 \)~\cite{michaud2013realization,chepiga2020floating}, beyond which it enters a fully dimerized phase. This transition belongs to the SU(2)\(_3\) WZW universality class with central charge \( c = 9/5 \)~\cite{michaud2013realization}.

We next examine the features of the DSFs in the dimerized phase using the SMA (see Item~\ref{item:domain1} in Subsection~\ref{SMAsection}), in order to gain a deeper understanding of the spectral properties of the $J_1$–$J_3$ model for spin-$1/2$, spin-1, and spin-$3/2$ chains.

\begin{figure*}[htbp]
    \centering
    \includegraphics[width=0.9\textwidth]{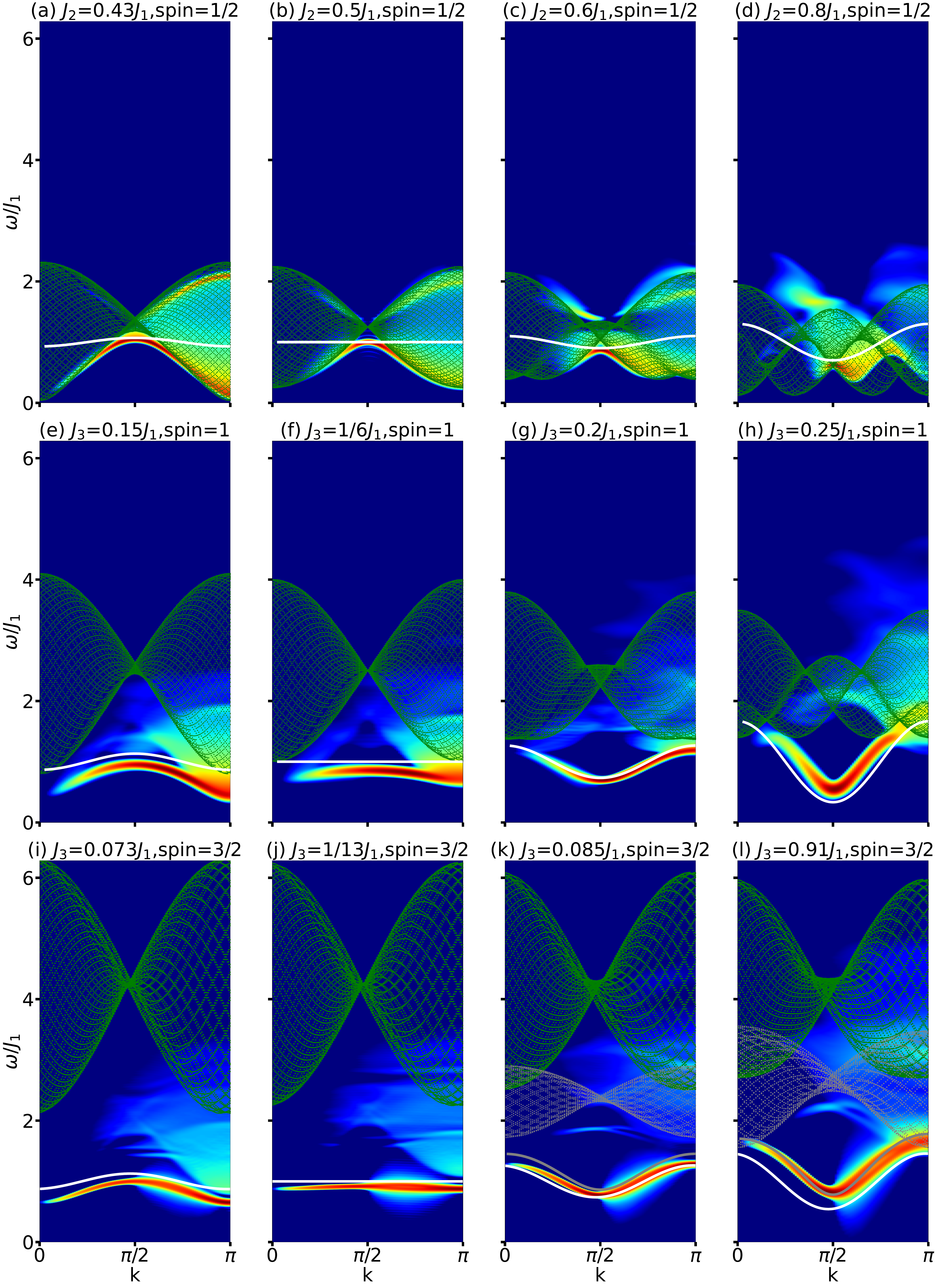} 
    \caption{Green shaded regions overlaid on the DSF indicate the two-domain-wall continuum computed using the SMA, for the fully dimerized phase of the $J_1$--$J_3$ model. Solid white lines represent the magnon dispersion obtained from SMA. (a–d) show results for the spin-$1/2$ chain, (e–h) for the spin-1 chain, and (i–l) for the spin-$3/2$ chain. In addition, panels (k) and (l) include magnon dispersions (gray lines) obtained from SMA applied to the DMRG ground state, along with the corresponding two-magnon continua (gray shaded regions).}
    \label{fig:smaj1j3cont}
\end{figure*}

\begin{figure*}[htbp]
    \centering
    \includegraphics[width=.7\textwidth]{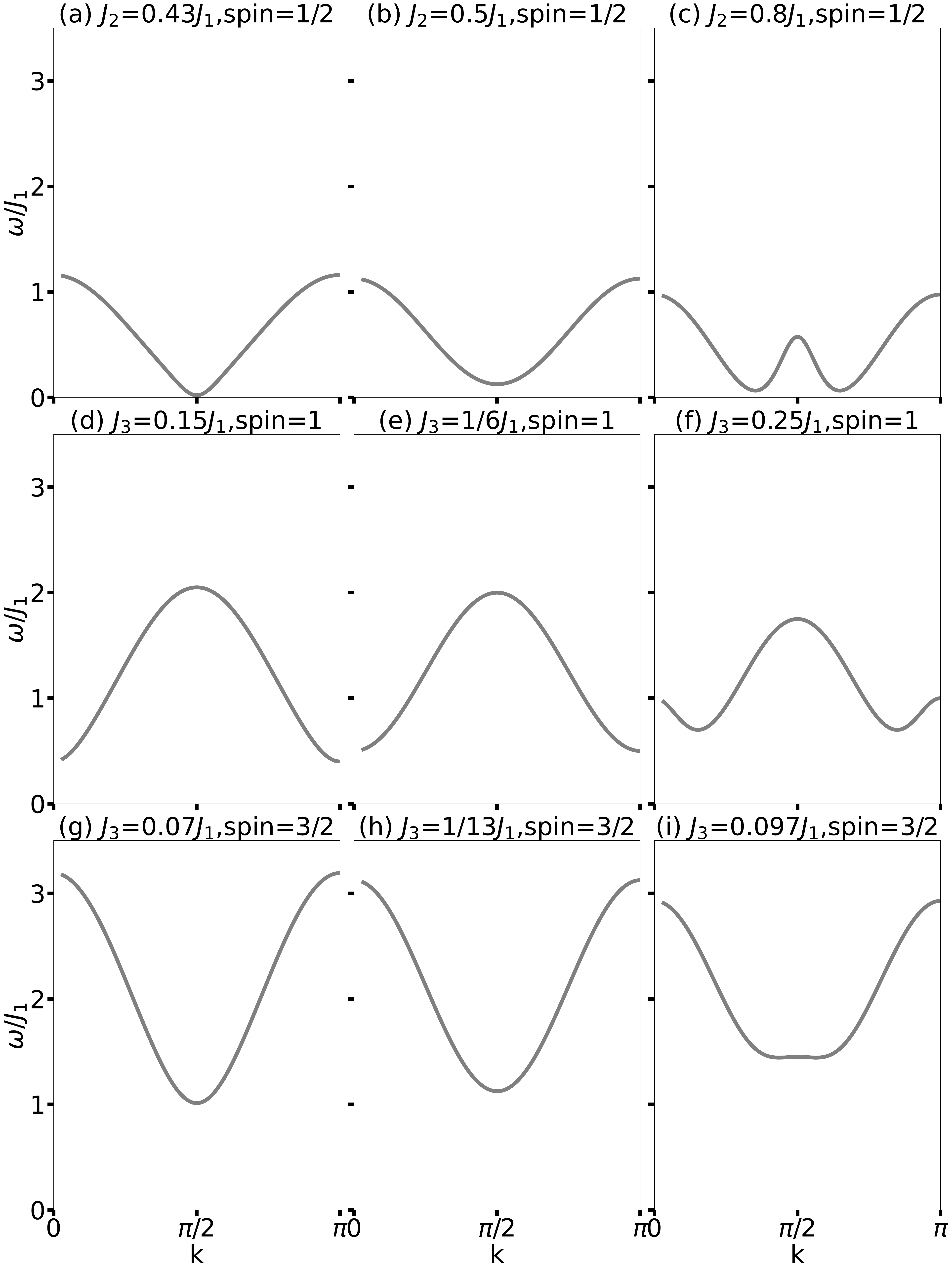} 
    \caption{Dispersion relations of domain-wall excitations in the $J_1$–$J_3$ Heisenberg chain for spin-1/2, spin-1, and spin-3/2. (a–c) show the domain-wall dispersion computed for the spin-1/2 chain, (d–f) for the spin-1 chain, and (g–i) present domain-wall dispersions for the spin-3/2 chain.
    }
    \label{fig:smaj1j3disp}
\end{figure*}

The spinon dispersion and continuum results for the spin-1/2 chain are well established in the literature \cite{lavarelo2014spinon} and we present them together with the new results of the spin-1 and spin-3/2 chains for completeness. The computed DSF spectra, overlaid with the two-domain wall continua (shown in green) and magnon dispersions obtained using the SMA, are shown in Fig.~\ref{fig:smaj1j3cont}. It provides a framework for interpreting the observed spectral features.

For the spin-1/2 chain [Figs.~\ref{fig:smaj1j3cont}(a–d)], the magnon dispersion (solid white lines) lies entirely inside the two spinon continuum, except around \(k=\pi/2\) near the MG point, where it briefly emerges. This aligns with the known presence of spinon bound states at \(k=\pi/2\) at the MG point \cite{lavarelo2014spinon} and points to the interpretation of this magnon as a bound state of spinons.
For the spin-1 chain [Figs.~\ref{fig:smaj1j3cont}(e–h)], the magnon dispersion lies entirely below the two domain wall continuum over most of the Brillouin zone. At its respective MG point, the magnon mode just touches the lower edge of the continuum near $k = 0$ and $k = \pi$. The lowest-energy excitation observed in the DSF lies close in energy to this dispersion and follows the same momentum dependence, strongly suggesting a common magnonic origin. For spin-3/2, there is a large separation between the magnon dispersion and the lowest edge of the domain-wall continuum [Figs.~\ref{fig:smaj1j3cont}(i–l)]. The magnon dispersion, obtained via SMA, closely tracks the lowest-energy mode of the DSF, affirming its magnon-like origin rather than the domain-wall character. Therefore, examining the evolution of the DSF in the fully dimerized phase across different spin values using SMA-derived excitations, one finds that as the magnitude of spin increases, the magnon mode shifts further from the domain-wall continuum. This suggests that domain-wall excitations are more energetically favorable in the spin-\(1/2\) case, while for higher spins such as 1 and \(3/2\), the system increasingly favors magnon mode excitations over domain-wall excitations.

Key conclusions from the DSF spectra and the SMA analysis are as follows:
(a) spinon fractionalization dominates in the spin-1/2 case, leading to a deconfined excitation continuum.
(b) magnon excitations persist for the spin-1 chain for all $J_3$ (except near the critical point), signaling the stability of the Haldane phase for the small $J_3$, and the dimerized state for the large $J_3$. For spin-3/2 chain, the spectrum is gapless until $J_3\approx 0.064 J_1$ beyond which magnon excitations are prominent. A common feature in both spin-1 and spin-3/2 chains is the presence of nearly dispersionless low-energy excitation modes near the MG points, a feature absent in the spin-1/2 system. Moreover, this feature exhibits a spin-dependent behavior, becoming progressively less dispersive with increasing spin.
(c) Domain walls arise as probable new higher-energy excitations above the magnon mode in the dimerized phase. 

A remark is in order regarding the nature of the excitation continuum in the dimerized phase. While domain-wall excitations provide a natural explanation for part of the observed continuum in the DSF, they do not account for all spectral features. Notably, remnants of the continuum that characterize the critical phase persist into the dimerized regime, including at the MG point. These remnants are visible in the DSF for both spin-1 and spin-3/2 chains and consistently appear near \( k = \pi \) and at energies around or above \( \omega/J_1 \approx 1 \), independent of the spin value. However, it is not possible to come up with a relevant VBS state and perform SMA calculations for this type of excitation. 

In addition, we note the presence of a distinct continuum of excitations in the spin-\( 3/2 \) chain in the DSF spectra at \( J_3/J_1 = 0.085 \) and \( 0.097 \) [Figs.~\ref{fig:smaj1j3cont}(k,l)], centered around \( k = \pi \) and \( \omega/J_1 \approx 2\text{--}4 \). This feature likely originates from a two-magnon continuum. The gray shaded regions in the figures correspond to the two-magnon continuum, constructed by combining momenta and energies of two noninteracting magnons using the dispersion obtained from the single-mode approximation (SMA) applied to the DMRG ground state, as defined in Eqs.~(\ref{Eq_sma_state}) and~(\ref{Eq_sma_defn}), with \( \Omega_j = S^z_j \). The corresponding SMA-derived magnon dispersion is shown as gray lines.

To gain further insight into the nature of excitations in the \(J_1-J_3\) model, we analyze the dispersion relations of a single domain wall in different spin systems in Fig.~\ref{fig:smaj1j3disp}. These single domain-wall dispersions are used to compute the spinon continua for spin-1/2 chains and the domain-wall continua for spin-1 and spin-3/2 chains and have been discussed above. An important observation is the location of the dispersion minima, which varies across different spin values. In the spin-1/2 chain [Figs.~\ref{fig:smaj1j3disp}(a–c)], the spinon dispersion exhibits a minimum at \(k=\pi/2\). For the spin-1 chain [Figs.~\ref{fig:smaj1j3disp}(d–f)], the domain-wall dispersion exhibits minima at \(k=0\) and \(k=\pi\). Finally, in the spin-3/2 chain [Figs.~\ref{fig:smaj1j3disp}(g–i)], the domain-wall dispersion has a minimum at \(k=\pi/2\), similar to the spin-1/2 case. This suggests that the fundamental excitations in the spin-1/2 and spin-3/2 systems differ qualitatively from those in the spin-1 case, reinforcing the distinct nature of domain walls in dimerized spin-1 chains. We investigate this difference by reinterpreting the SMA-derived dispersion [Eq.~(\ref{Eq_sma_defn_trig})] in terms of a dispersing mode in an effective tight-binding model. We determine the effective tight-binding Hamiltonian in a non-orthogonal basis of the domain-wall states. The problem is recast as a generalized eigenvalue equation in which both the Hamiltonian and the overlap matrix are evaluated in this basis. This methodology is detailed in Refs.~\cite{lavarelo2014spinon,sharma2025bound,sharma3by2}.

At the MG point, we extract an effective tight-binding model for the $J_1$–$J_3$ chain, expressed as

\begin{align}
\omega(k) = \gamma_0 + \sum_{n=1}^{N/2-1} 2\gamma_n \cos(kn),
\label{eq:tight_binding_dispersion}
\end{align}
where \( \gamma_0 \) is the onsite energy, and \( \gamma_n = \bra{\Omega_{N/2}} \tilde{H} \ket{\Omega_{N/2 + n}} \) are the effective hopping amplitudes in the tight-binding description. Here, \( \tilde{H} \) denotes the projected Hamiltonian, and \( \ket{\Omega_r} \) is the VBS state with domain walls located at position \( r \). The index \( n \) corresponds to the separation between domain wall positions. We note for clarity that this is not an approximation but an exact reformulation of Eq.~(\ref{Eq_sma_defn_trig}). 

We find that the second-neighbor hopping amplitude is negative for the spin-1 chain, while it is positive for the spin-1/2 and spin-3/2 chains. This sign difference underlies the qualitative difference in the dispersion minima between spin-1 and the other two cases. Additionally, as \(J_3\) increases, we observe a splitting of the dispersion at these minima in all three cases.

The values of \( \gamma_0 \) and \( \gamma_2 \) for each spin are tabulated below. All other \( \gamma_n \) are zero.

\begin{table}[ht]
\centering
\caption{Effective tight-binding hopping amplitudes \( \gamma_n \) extracted from SMA domain-wall dispersions for various spin values at the MG point.}
\begin{tabular}{ccc}
\toprule
Spin & \( \gamma_0 \) & \( \gamma_2 \) \\
\midrule
$\frac{1}{2}$ & 0.625 & 0.25 \\
1             & 1.25  & -0.375 \\
$\frac{3}{2}$ & 2.125 & 0.5 \\
\bottomrule
\end{tabular}
\label{tab:gammas}
\end{table}

From these effective parameters, we conjecture the following compact analytical expression that reproduces the SMA domain-wall dispersion at the MG point:

\begin{equation}
\omega(k) = \frac{1}{4} + \frac{S(S+1)}{2} + \frac{(-1)^{2S+1}(2S+1)}{4} \cos(2k)
\end{equation}
\vspace{15pt}

Notably, for \( S = 1/2 \), this expression reduces to the exact domain-wall dispersion obtained in Ref.~\cite{shastry1981excitation}. We have further verified that the hopping amplitudes presented in Table~\ref{tab:gammas} remain valid not only for the \( J_1 \)–\( J_3 \) model (\( J_2 = 0 \)), but also along the full exactly dimerized lines (depicted in Fig.\ref{phasediags}) described by the analytical expression in Ref.~\cite{wang2013dimerizations}, which includes cases with finite \( J_2 \).

\section{DSF along the phase transition from Haldane phase to dimerized phase in the spin-1 chain}
\label{spin1_J1_J2_J3}

\begin{figure*}[htbp]
    \centering
    \includegraphics[width=\textwidth]{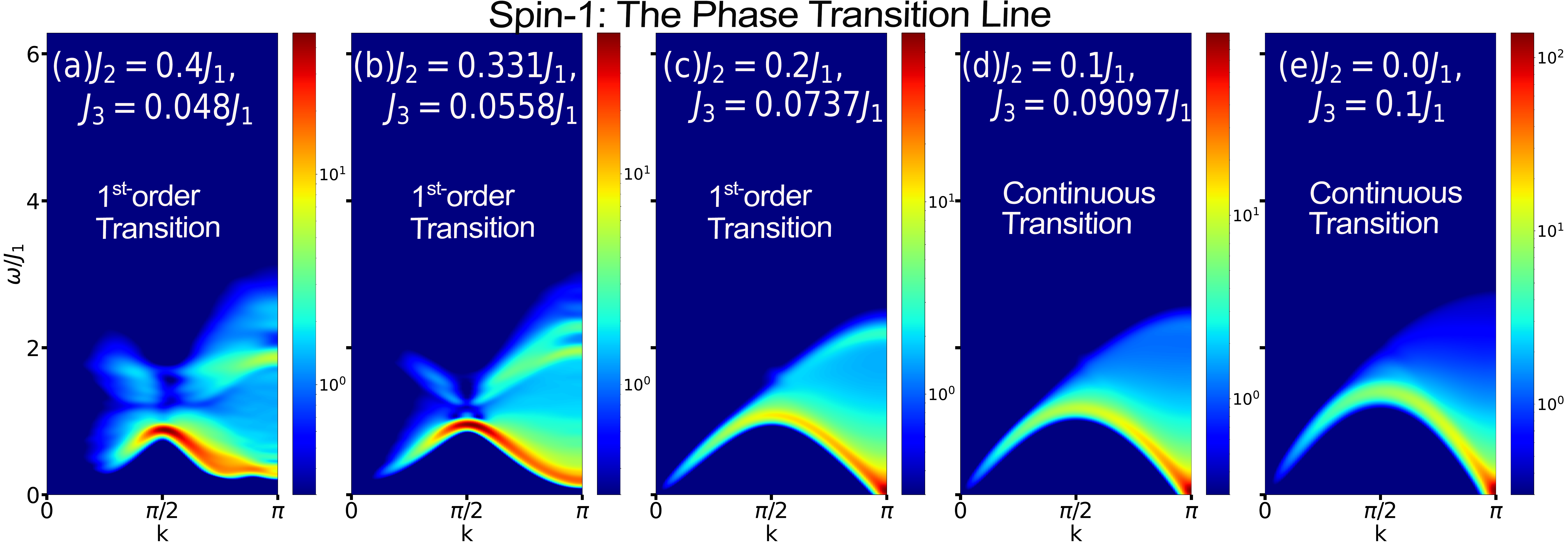} 
    \caption{DSF $S(q,\omega)$ along the Haldane-to-dimerized transition line in the spin-1 $J_1$–$J_2$–$J_3$ chain. Fig (a–c) lie in the first-order regime, while fig (d–e) lie in the second-order regime.
    }
    \label{fig:spin1trans}
\end{figure*}

\begin{figure*}[htbp]
    \centering
    \subfigure[]{
        \includegraphics[width=1\textwidth]{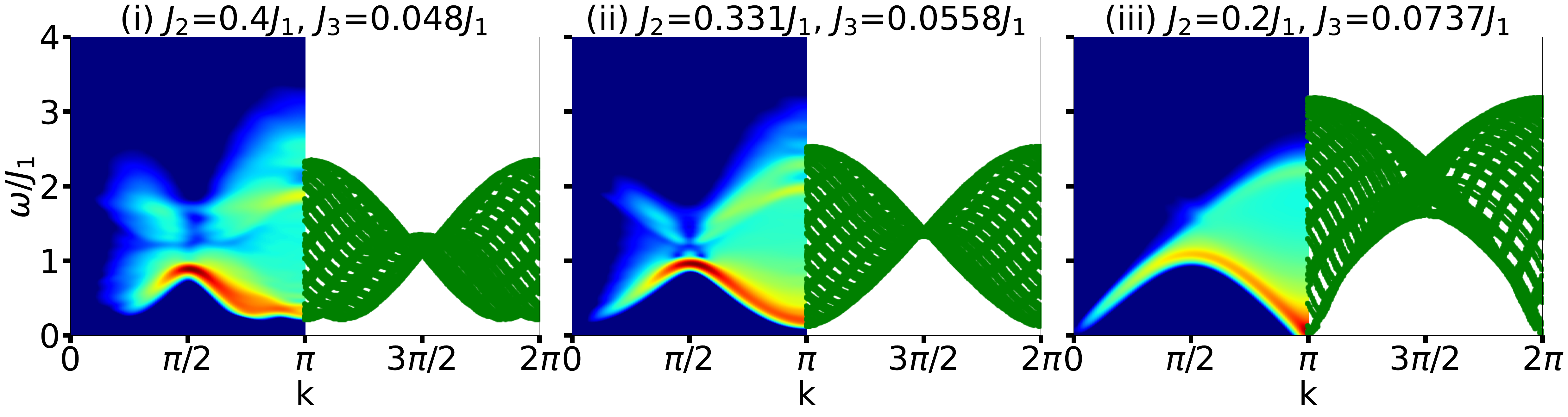}
        \label{}
    } \\
    \subfigure[]{
        \includegraphics[width=0.975\textwidth]{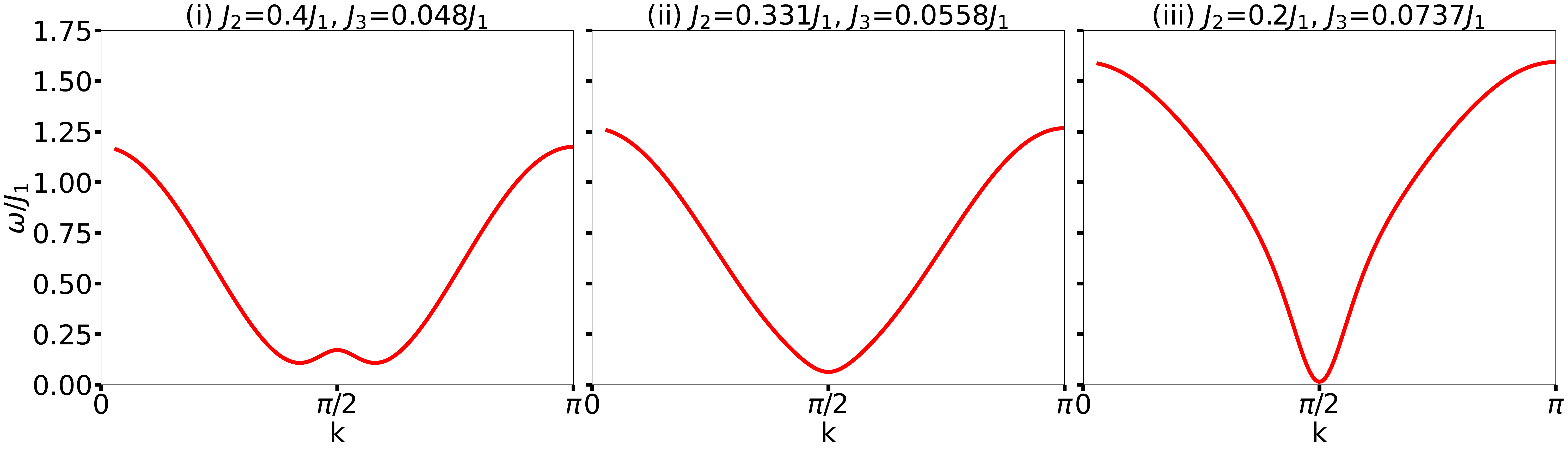}
        \label{}
    }
        \caption{(a) Green shaded regions compared with the DSF indicate the two-domain-wall continuum along the line of first-order transitions (see Fig.~\ref{phasediags}a), computed using the SMA. Panels (b)(i), (ii), (iii) display the corresponding dispersion relations of domain-wall excitations associated with the shaded continua in the $J_1$–$J_2$–$J_3$ Heisenberg chain for spin-1, along the line of first-order transitions between Haldane and dimerised phases. The simulations were performed on systems of 120 sites.}
    \label{fig:transvbs1}
\end{figure*}

In this section, we analyze the DSF $S^{zz}(q,\omega)$ along the transition line between the Haldane phase and the dimerized phase in the spin-1 $J_1$–$J_2$–$J_3$ chain. For a fixed $J_2$ coupling, the model with the smaller $J_3$ couplings lies in the Haldane phase and has a translation-invariant ground state whereas across the transition line, the model with larger $J_3$ couplings lies in a dimerized phase characterized by ground state with broken translational symmetry. The nature of the phase transition changes along this line: it is second order near $J_2 = 0$ and becomes first order as $J_2$ increases, consistent with the field theory arguments discussed in Ref.~\cite{chepiga2016dimerization,chepiga2016spontaneous}.

The evolution of the DSF spectra along this line is shown in Fig.~\ref{fig:spin1trans}. Our results reveal distinct features in DSF that reflect the underlying change in the phase transition character. Fig.~\ref{fig:spin1trans}(a)–(c) depict points along the first-order transition regime, while Fig.~\ref{fig:spin1trans} (d)–(e) correspond to the second-order regime near \(J_2 = 0\), as established by the parameter space in the phase diagram shown in Fig.~\ref{phasediags}(a). Along the second-order segment of the transition (i.e., for small $J_2$) [Fig.~\ref{fig:spin1trans}(d–e)], the spectrum exhibits soft modes, consistent with critical behavior governed by the SU(2)$_2$ WZW universality class. There is a low-energy spectral weight concentrated near the commensurate wavevector $q = \pi$, with a broad continuum extending toward higher energies. In contrast, along the line of first-order transitions (for larger $J_2$)  [Fig.~\ref{fig:spin1trans}(a–c)], the DSF remains gapped across the transition, with well-defined excitations that do not soften. The finite gap reflects the energy cost of creating a pair of domain walls between regions of Haldane and dimerized order. It is difficult to determine, based solely on the low-energy part of the spectrum, whether the DSF in Fig.~\ref{fig:spin1trans}(c) is truly gapless. Ref.~\cite{chepiga2016dimerization} notes that the gap opens exponentially slowly just beyond the endpoint of the SU(2)\(_2\) WZW critical line and may be visible only in much larger systems than those studied here. 

At a point along the first-order segment, a domain wall separating the Haldane VBS state and the dimerized spin-$1$ VBS state necessarily has a free spin-1/2 degree of freedom and is a fractional excitation (spinon), as shown in Fig.\ref{fig:domainwalldis}(a). Since the ground states in the adjacent phases are degenerate at the first-order phase transition, these pairs of spinons are deconfined and result in a continuum in the DSF [see Fig.~\ref{fig:spin1trans}(a–c)]. In the DSF, we additionally note that the extent of the continuum is consistent with the SMA calculations, as shown in Fig.~\ref{fig:transvbs1}(a). The single spinon dispersion from SMA which is used to compute the spinon continuum is shown in Fig.~\ref{fig:transvbs1}(b). The good agreement of the SMA-determined spinon continuum with the DSF spectra is strong evidence that the low-energy excitations at the line of first-order transitions are deconfined spinons.

An important aspect of the transition line is the change in the nature of spin correlations. For larger $J_2$ values, the spin-spin correlations exhibit incommensurate oscillations, a hallmark of frustrated interactions and competing length scales. As one moves along the transition line toward smaller $J_2$, these correlations gradually become commensurate \cite{chepiga2016dimerization}. This commensurate-incommensurate (C-IC) crossover is mirrored in the DSF [see Fig.~\ref{fig:spin1trans}(a–c)] by a change in the momentum of the energy minima of the spinon continuum, from incommensurate wavevectors toward $q = \pi$. The C-IC crossover is also consistent in the individual spinon dispersion determined from SMA [Fig.~\ref{fig:transvbs1}(b)] as its energy minima in momentum shifts from incommensurate values to $\pi/2$. See also Item~\ref{item:domain2} in Subsection~\ref{SMAsection} for a discussion of domain-wall excitations in this regime.

\section{Spinon confinement and the disintegration of the continuum across first-order transitions in spin-$1$ and spin-$3/2$ chains}
\label{spinon_confinement}

\begin{figure*}[htbp]
    \centering
    \subfigure[Spin-1]{
        \includegraphics[width=1\textwidth]{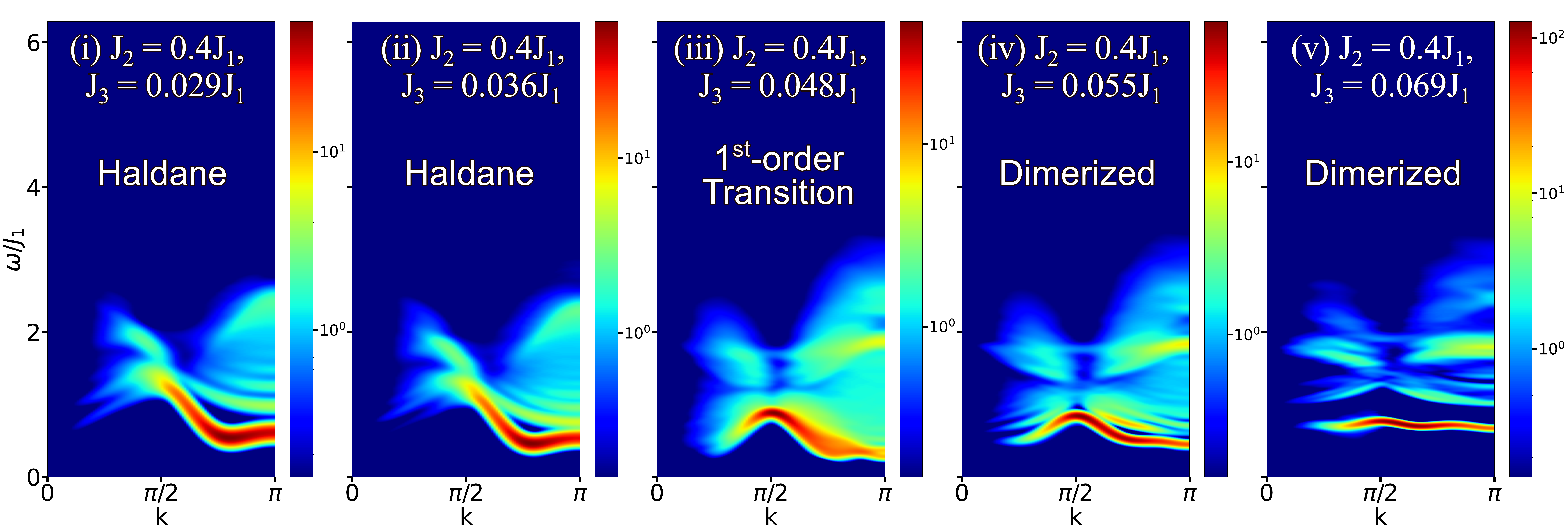}
        \label{}
    } \\
    \subfigure[Spin-3/2]{
        \includegraphics[width=1\textwidth]{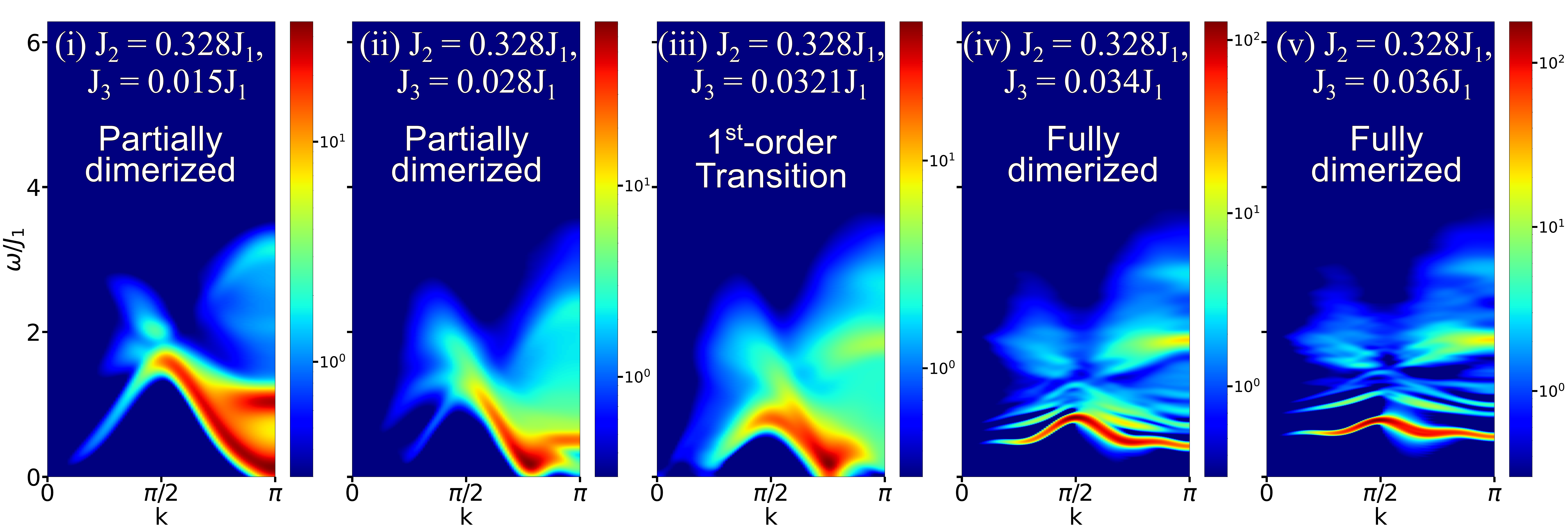}
        \label{}
    }
        \caption{DSF across the line of first-order transitions in spin-$1$ and spin-$3/2$ $J_1$–$J_2$–$J_3$ chains. Figures a(i–v) show the evolution of the DSF for the spin-1 chain at fixed $J_2 = 0.4J_1$ (cut C1, see Fig.~\ref{phasediags}a) as $J_3/J_1$ is tuned across the first-order transition between the Haldane and dimerized phases. Figure a(iii) is situated at the transition point. Figures b(i–v) depict the corresponding evolution for the spin-$3/2$ chain at $J_2 = 0.328J_1$ (cut C2, see Fig.~\ref{phasediags}b), traversing the first-order transition from a partially dimerized to a fully dimerized phase. Figure b(iii) is at the transition point. System sizes used in simulations are 120 sites for spin-$1$ chains, and 150 sites for spin-$3/2$ chains. 
        }
    \label{fig:dsftrans}
\end{figure*}

\begin{figure*}[htbp]
    \centering
    
        \includegraphics[width=1\textwidth]{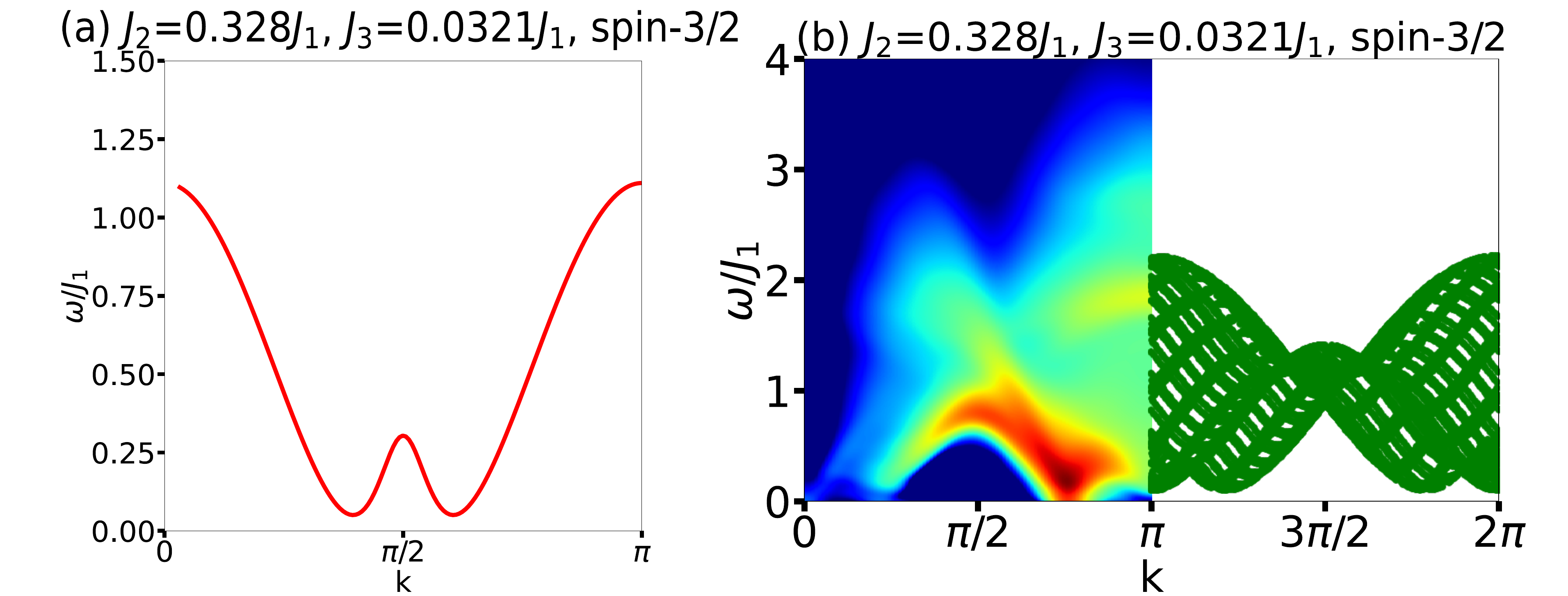}
        \label{}
        \caption{(a) shows the domain-wall dispersion relation obtained using the SMA at a representative first-order transition point in the $J_1$–$J_2$–$J_3$ Heisenberg chain for spin-$3/2$. This dispersion corresponds to spin-$1/2$ domain wall separating the partially and fully dimerized states. (b) displays the resulting two-domain-wall continuum, shown as green shaded region, compared with the DSF. This continuum is constructed by combining non-interacting spinon excitations derived from the SMA.}
    \label{fig:3by2_dispersion}
\end{figure*}

In this section, we examine how spinons evolve across the lines of first-order transitions in the spin-1 and spin-3/2 $J_1$–$J_2$–$J_3$ chains \cite{chepiga2016dimerization,chepiga2020floating}. For spin-$1$ chain, the first-order transition is between the Haldane phase and the dimerized phase while for spin-$3/2$ chain, the first-order transition is between the partially dimerized and fully dimerized phases. Despite the distinct nature of the adjacent phases, both systems reveal a common dynamical feature: the formation of domain walls carrying spin-$1/2$ at the first-order transition, and their confinement into discrete bound states far away from the transition (see Fig.~\ref{fig:dsftrans}).

For the spin-1 chain, we focus on the first-order transition at $J_2 = 0.4 J_1$ (cut C1, see Fig.~\ref{phasediags}a), varying $J_3$ across the transition point. The DSF shows the evolution of the spectra with broad continuum of deconfined spinons at the transition point and as one moves away from it, the spectra proliferate into discrete bound states of spinons [see Fig.~\ref{fig:dsftrans}(a)]. This is because the spinons experience an effective confining potential arising due to the energy cost of a higher-energy state enclosed between the domain walls when they are separated in the chain.

A parallel scenario unfolds in the spin-$3/2$ chain, where we concentrate on the first-order transition at $J_2 = 0.328 J_1$ (cut C2, see Fig.~\ref{phasediags}b) across multiple $J_3$ values. The DSF for this case is shown in Fig.~\ref{fig:dsftrans}(b). Here, the transition is between the partially dimerized phase, characterized by alternating strong and weak bonds, and a fully dimerized phase with maximal singlet coverage. At the transition point [Fig.~\ref{fig:dsftrans}b(iii)], the DSF reveals a gapped continuum consistent with deconfined spinons as expected. As in the spin-1 chain, the domain walls are fractional excitations of spin-1/2, and there is a small energy cost for creating  pairs of such spinons at the transition point which is reflected in the presence of a spectral gap in the DSF. Similar to the spin-$1$ chain, as one moves away from the transition the DSF spectra in the fully dimerized phase [Fig.~\ref{fig:dsftrans}b(iv-v)] bunches into bound states of spinons because of the confining potential experienced by the spinons.

To further confirm the nature of the excitations observed at the transition point, we perform SMA calculations  (see Item~\ref{item:domain2} in Subsection~\ref{SMAsection}) to extract the spinon dispersion in both spin-1 and spin-3/2 chains. Since the spinons are deconfined at the transition, we construct the corresponding two-spinon continua by summing over the momenta and energies of the spinons. The resulting continua are shown as shaded green regions in Fig.~\ref{fig:transvbs1}a(i) for the spin-1 chain, and in Fig.~\ref{fig:3by2_dispersion}(b) for the spin-3/2 chain. The corresponding spinon dispersion for the spin-3/2 chain is shown in Fig.~\ref{fig:3by2_dispersion}(a). Remarkably, the shape, bandwidth, and location of the minima on the momentum axis of the green shaded regions agree closely with those of the DSF. The close match between the SMA predictions and the DSF spectra offers compelling confirmation that the low-energy excitations at the first-order transition in both spin-1 and spin-3/2 chains are well described by fractionalized domain walls propagating in a background of competing VBS states.

\section{Summary and Conclusions}
\label{summary}

In this work, we perform a comprehensive numerical investigation of the DSF for the frustrated $J_1$–$J_2$–$J_3$ Heisenberg spin chains with spin magnitudes $S = 1/2$, $1$ and $3/2$  using time-dependent DMRG and the SMA. The nature of elementary excitations in the phases and the evolution of these excitations across phase transitions were discussed.

For the $J_1$--$J_3$ model (i.e., in the absence of $J_2$), we observe a clear spin-dependent evolution of the excitation spectra. In the spin-$1/2$ chain, the low-energy dynamics is dominated by fractionalized spinon continua, reflecting strong quantum fluctuations. In contrast, the spin-1 and spin-$3/2$ chains exhibit sharp, well-defined magnon modes as the dominant low-energy excitations in the dimerized phase. Additionally, domain-wall excitations emerge as probable higher-energy modes situated above the magnon dispersion. Notably, the relative position of the SMA-derived magnon and domain-wall continua is strongly spin dependent. For spin-1, the magnon mode lies below the domain-wall continuum across most of the Brillouin zone, touching its lower boundary at the MG point, while for spin-$3/2$ this separation becomes even more pronounced. This trend suggests that domain-wall excitations are more energetically favorable in the spin-$1/2$ chain, whereas for higher spins, the system increasingly stabilizes magnon mode excitations over domain-wall excitations.

Along the Haldane-to-dimerized transition in the spin-1 chain, the DSF clearly reflected the changing nature of the transition from second-order to first-order as frustration parameters varied. The second-order segment exhibited gapless critical excitations characteristic of a continuous transition governed by SU(2)$_2$ WZW universality \cite{chepiga2016spontaneous,chepiga2016dimerization,affleck1987critical,affleck1986exact}, while the first-order segment displayed well-defined gapped excitations arising from domain walls between competing Haldane and dimerized phases. Notably, despite this change in the order of the transition, the excitation spectrum remains qualitatively similar throughout, in the sense that a continuous band of excitations persists along the entire transition line.

Furthermore, across first-order transitions, such as between the Haldane and dimerized phases in spin-1 chains, and between partially and fully dimerized phases in spin-3/2 chains, we find a universal spinon confinement mechanism. At the transition, spinon domain walls are deconfined, forming broad excitation continua. However, upon entering the fully dimerized phases, confinement sets in rapidly which results in continua proliferating into discrete bound states of spinons. This dynamical signature of confinement was clearly visible in the DSF, underscoring the universal nature of fractionalization and confinement phenomena in frustrated spin chains \cite{vanderstraeten2020spinon,sharma2025bound}.

Nowadays, the DSF can be obtained numerically, but its structure is often very complex. This work demonstrates that the SMA, applied to different types of excitations, is a powerful tool for identifying the underlying degrees of freedom from which the spectral function can be reconstructed. We believe that this approach could be useful in understanding the DSF of many other one-dimensional quantum models.

\begin{acknowledgments}
A.S. acknowledges support from the Swiss Government Excellence Scholarship (FCS Grant No. 2021.0414). This work was supported by the Swiss National Science Foundation under Grant No. 212082 and 188648. The calculations have been performed using the facilities of the Scientific IT and Application Support Center of EPFL.
\end{acknowledgments}

\bibliography{references}     

\appendix

\section*{APPENDIX}

\section{Incommensurability and Momentum Crossover in the Spin-1 $J_1$–$J_3$ Chain}
\label{appendix:dsf_comm}

In this appendix, we report supplementary results on the DSF of the spin-1 $J_1$–$J_3$ chain in the parameter range between the generalized MG point ($J_3/J_1 = 1/6$) and $J_3/J_1 = 0.2$. Motivated by previous work~\cite{chepiga2016spontaneous}, we performed calculations to investigate the possible onset of incommensurability in the DSF in this regime. For the present simulations, the maximum evolution time was $t_{\mathrm{max}} = 240/J_1$ and the length of the lattice was at 120 sites.

\begin{figure*}[t]
    \centering
    \includegraphics[width=\linewidth]{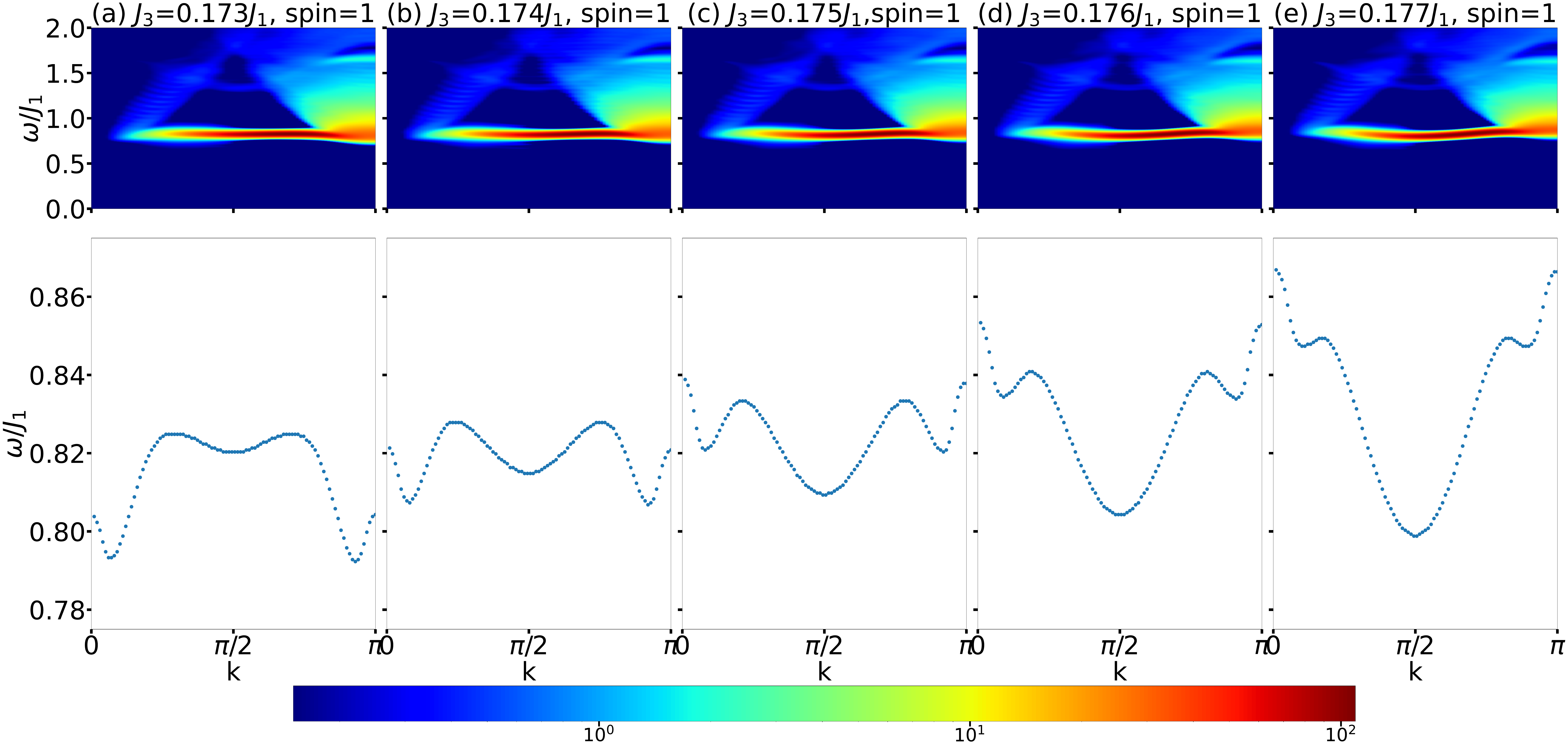}
    \caption{Top panels: dynamical structure factor $S^{zz}(k,\omega)$ of the spin-1 $J_1$–$J_3$ chain for $J_3/J_1 = 0.173$–$0.177$. Bottom panels: corresponding magnon dispersions extracted from the DSF. The momentum of the lowest-energy mode changes from incommensurate to $k = \pi/2$ between $J_3/J_1 \approx 0.174–0.175$.}
    \label{fig:j1j3_crossover}
\end{figure*}

Fig.~\ref{fig:j1j3_crossover} shows DSF plots (top panels) and the corresponding dispersions extracted following the maximum intensity contours (bottom panels) for five closely spaced values of $J_3/J_1$ in the range $0.173$–$0.177$. The dispersions show that the lowest-energy mode in the DSF, located at $k = \pi$ for $J_3$ at the disorder point ($J_3/J_1 = 1/6$, Fig.~\ref{fig:dsfsj1j3spin1and3by2}h), is already incommensurate at $J_3 \approx 0.173 J_1$ (Fig.~\ref{fig:j1j3_crossover}a). This is consistent with earlier findings that the disorder point lies at $J_3/J_1 = 1/6$, while the Lifshitz point, where the static structure factor becomes incommensurate, occurs at $J_3 \approx 0.185 J_1$~\cite{chepiga2016spontaneous}. The incommensurability in the DSF is thus observed in the intermediate region between these two points.

A more detailed inspection of the magnon dispersion extracted from the DSF shows that the momentum $k_{\mathrm{min}}$ of the lowest-energy mode changes abruptly from an incommensurate value to $k = \pi/2$ in the range $J_3/J_1 \approx 0.174 - 0.175$ [Fig.~\ref{fig:j1j3_crossover}(a–e)]. 



 Our numerical results suggest a crossover between two regimes: one with an incommensurate minimum evolving from $k = \pi$, and another with the minimum locked at $k = \pi/2$ at and beyond $J_3/J_1 \approx 0.175$. 

\end{document}